\documentclass[aps, prl, reprint,graphicx,groupedaddress,amsmath,amssymb]{revtex4-1}
\usepackage{dcolumn}
\usepackage{bm}
\usepackage{natbib}
\usepackage{bibentry}
\usepackage[usenames,dvipsnames] {color}
\usepackage[table]{xcolor}
\usepackage{graphicx}
\usepackage{epstopdf}
\usepackage{amssymb}
\usepackage{amsmath}
\usepackage{dcolumn}
\usepackage{subfigure}
\usepackage{titlesec}

\DeclareGraphicsExtensions{.eps,.png,.pdf}

\usepackage{xr-hyper}
\usepackage{hyperref}
\hypersetup{
colorlinks   = true, 
urlcolor     = blue, 
linkcolor    = blue, 
citecolor   = blue 
}
\usepackage{setspace}
\usepackage[font=singlespacing,font=footnotesize, justification=RaggedRight]{caption}
\usepackage{enumitem}
\usepackage{fancybox}
\usepackage{comment}
\usepackage{url}
\usepackage{here}
\setlist{nolistsep}
\usepackage{multimedia}
\usepackage{soul}
\setstcolor{red}
\DeclareMathAlphabet      {\mathbf}{OT1}{cmr}{bx}{n}

\newcommand{\printfnsymbol}[1]{%
\textsuperscript{\@fnsymbol{#1}}%
}
\makeatother

\newcommand{\RomanNumeralCaps}[1]
{\MakeUppercase{\romannumeral #1}}

\definecolor{gold_metallic}{rgb}{0.83, 0.69, 0.22}
\definecolor{frenchlilac}{rgb}{0.53, 0.38, 0.56}
\definecolor{indigo}{rgb}{0.0, 0.25, 0.42}
\definecolor{wildwatermelon}{rgb}{0.99, 0.42, 0.52}
\definecolor{wildstrawberry}{rgb}{1.0, 0.26, 0.64}
\definecolor{cadetblue}{rgb}{0.37, 0.62, 0.63}
\definecolor{caribbeangreen}{rgb}{0.0, 0.8, 0.6}

\graphicspath{{../Figures/}{../SI/}}

\begin{document}

\title{Defect Spirograph: Dynamical Behavior of Defects in Spatially
Patterned Active Nematics}

\author{Ali Mozaffari$^{1}$}
\thanks{equal contribution}
\author{Rui Zhang$^{1,2}$}
\thanks{equal contribution}
\author{Noe Atzin$^{1}$}
\author{Juan J. de Pablo$^{1,3}$}
\email{depablo@uchicago.edu}

\affiliation{
 $^1$Pritzker School of Molecular Engineering, The University of Chicago, Chicago, Illinois
 60637, USA \\
 $^2$Department of Physics, Hong Kong University of Science and
  Technology, Clear Water Bay, Kowloon, Hong Kong\\
 $^3$Center for Molecular Engineering, Argonne National Laboratory, Lemont, Illinois
 60439, USA
}

\date{\today}

\begin{abstract}
  Topological defects in active liquid crystals can be confined by introducing
  gradients of activity. Here, we examine the dynamical behavior of two defects confined by
  a sharp gradient of activity that separates an active circular region and a surrounding passive nematic
  material.
  Continuum simulations are used to explain how the interplay among energy injection into
  the system, hydrodynamic interactions, and frictional forces governs the dynamics
  of topologically required self-propelling $+1/2$ defects.
  Our findings are rationalized in terms of a phase diagram for the
  dynamical response of defects in terms of activity and frictional damping strength.
  Different regions of the
  underlying phase diagram correspond to distinct dynamical modes,
  namely immobile defects
 (ID), steady rotation of defects (SR), bouncing defects (TB), bouncing-cruising defects
 (BC), dancing defects (DA), and multiple defects with irregular dynamics (MD).
  These dynamic states raise the
  prospect of generating synchronized defect arrays for microfluidic applications.
\end{abstract}

\maketitle

Active nematics represent a class of non-equilibrium active systems with broken rotational
symmetry, where orientational ordering, elastic stresses, active stresses, and hydrodynamic
forces, interact to produce
unique structures and dynamical behaviors \cite{ramaswamy2010, marchetti2013, needleman2017,
doostmohammadi2018, shaebani2020, ignes2020, zhang2021auto}.
Active nematics have drawn considerable interest, and a general understanding of their behavior has begun to emerge from
observations across a wide range of experimental systems
that include \textit{in vitro}
assemblies of filamentous proteins \cite{sanchez2012, keber2014, kumar2018, rivas2020},
dense suspensions of elongated bacteria \cite{dunkel2013, li2019},
living nematics \cite{zhou2014},
and dense colonies of elongated cellular tissues \cite{duclos2017, kawaguchi2017, blanch2018, dell2018}.
These experiments have been accompanied by theoretical studies that have sought to explain or anticipate their properties
\cite{simha2002, narayan2007, marenduzzo2007, giomi2011,
giomi2013, giomi2015, hemingway2016, doostmohammadi2016b, shankar2018, norton2020, pearce2020, vafa2020, 
serra2021, mueller2021, mahault2021}.
In active nematics, active stresses lead to a constant generation and
annihilation of topological defects, giving rise to a loss of long-range
nematic ordering.
The flows associated with these defects are devoid of any spatiotemporal coherence;
even for initially ordered systems, regular dynamics and
pattern formation are transient \cite{martinez2019, sokolov2019}.
Attempts to control and design active nematics
are still in their infancy, and have been hindered by a limited understanding of the underlying coupling between structure
and dynamics. \par

Several reports have shown that geometrical confinement and the topological features
can be used to stabilize the
chaotic motion of active units and to harness their energy to generate well-defined flows
\cite{wioland2013, ravnik2013, keber2014, zhang2016, khoromskaia2017,
wu2017, ellis2018, guillamat2018, opathalage2019, hardouin2020, rajabi2021}. For example, theory predicts the
formation of a circulating pair of $+1/2$ defects under
circular confinement, with the bulk dynamics insensitive to the imposed anchoring condition 
\cite{norton2018}.
In unconfined systems, strategies such as assembling
extracts of cytoskeletal filaments and motor proteins on anisotropic soft
structured interfaces \cite{guillamat2016, guillamat2017}, or such as fixing the
underlying nematic in contact with a bacterial suspension through a photopatterning 
\cite{peng2016, turiv2020, lavrentovich2021}, have both been shown to tame the otherwise chaotic flows.
Past work has also sought to harness these flows by relying on frictional forces
\cite{wioland2013, thampi2014, doostmohammadi2016, pearce2019, thijssen2021}.
Very recently, the patterning of activity has been shown to be effective at
trapping and segregating topological charges, thereby providing a means to guide defect motion
\cite{zhang2021spatio, shankar2019, tang2021}.\par

\begin{figure}[!htb]
 \centering
 \includegraphics[width=0.98\columnwidth]{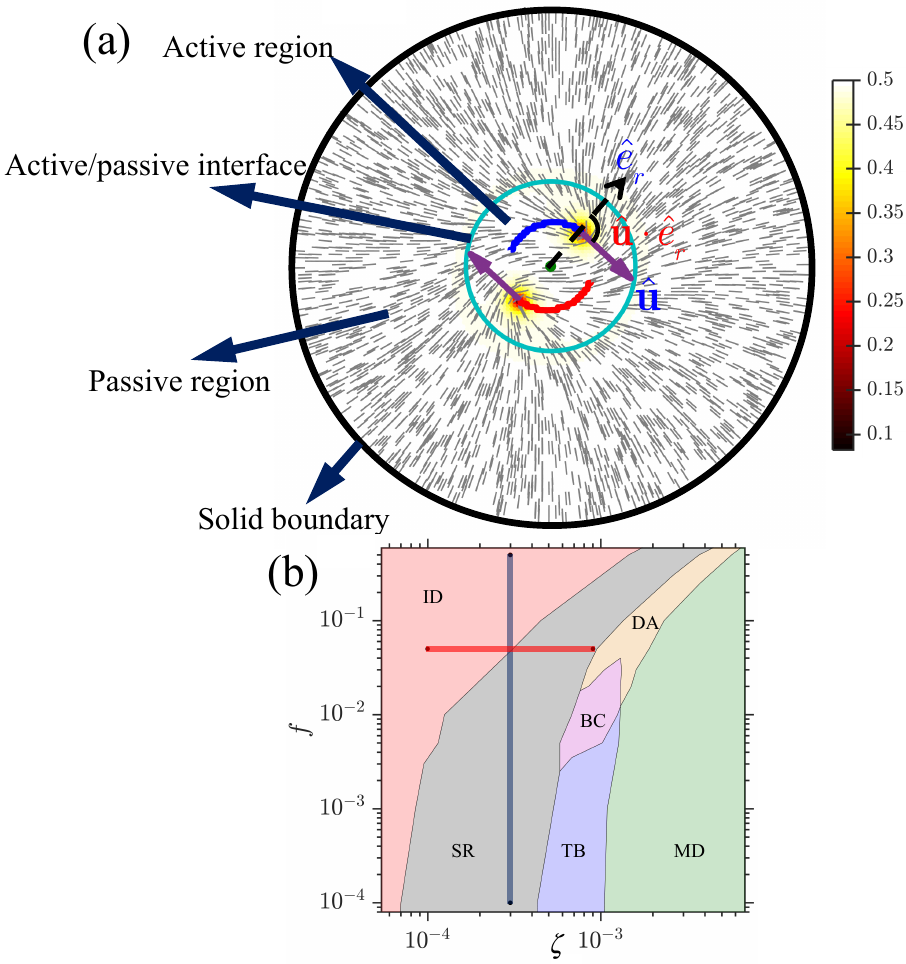}
 \caption{(a) An snapshot of defect trajectories within the activated
  region of nematic confined in a solid disk. Purple arrows show the
  orientation of $+1/2$ defects ($\hat{u}$).
  The color coding corresponds to the nematic ordering ($S$). (b) A phase diagram of defect 
  dynamics as a function of activity and friction coefficients.}
 \label{fig1}
\end{figure}

In this work, we rely on the relatively new concept of activity patterning or localization, along with the frictional
damping, to control defect dynamics in a quasi $2$D active nematic. Through this strategy, it becomes possible to
confine and further manipulate the dynamics of
two, like-charge $+1/2$ topological defects (Fig.~\ref{fig1}(b)).
These new dynamics should be contrasted with those
observed for two defects confined by standard hard walls, where momentum cannot be transferred across the boundary,
leading to different outcomes and a more limited set of behaviors.
\par

The spatiotemporal evolution of nematic tensorial order parameter $\bf{Q}$ and a flow field $\bf{u}$
follow the continuum equation of motion for an active nematic and are solved using a hybrid lattice Boltzmann method
\cite{marenduzzo2007, ravnik2013, zhang2016jcp, zhang2018}. The details of simulations and the full governing equations
are provided in the SI.
Simulations were performed on a $2$D lattice confined in a solid disk, with the homeotropic anchoring and
no-slip velocity field were enforced at the boundary. The system was initialized with the director field radially
oriented. The regions inside the circular domain within the disk were activated by applying
uniform extensile active stresses ${{\bm{\Pi}} ^a} =  - \zeta {\bf{Q}}$, with the activity strength $\zeta > 0$.
The domain outside the activated region is passive (Fig.~\ref{fig1}(a)). \par 

The Poincar\'e-Hopf theorem \cite{kamien2002} enforces the presence of topological defects of
total charge $+1$ in the current system. As such, the system will develop at least one pair
of $+1/2$ defects, which in the absence of activity experience repulsive
interactions that maximize their separation distance.
At the same time, the interface created by the spatially inhomogeneous activity profile repels the defects,
leading to their entrapment within the active region and the appearance of new patterns and spatiotemporal states
(see SI and Fig.~S1 for a discussion of the repulsive interaction between defects and soft activity interface). \par

At high values of activity, a transition occurs into a chaotic state, with a fluctuating number of
defects, irregular dynamics, continuous formation of bands and their unzipping, and an absence of
long-lived defects due to constant defect renovation.
As the friction increases, the threshold of activity required
for the transition to the MD state becomes larger.
The MD state develops when
the activity-induced length scale $\sqrt{L/\zeta}$ is smaller than the frictional screening
length $\sqrt{\eta/f}$ and the radius of the active circle $R$, where $L$ is the elastic constant,
$\eta$ is the medium viscosity, and $f$ the friction coefficient. \par

In passive systems, the free energy is minimized by the formation of two $+1/2$
defects close to the boundary \cite{vitelli2004}, whose separation distance is determined by the ratio of
the surface anchoring strength and the material's elastic constant. For very low values of activity or for high
frictional damping, the active system under consideration
behaves as a passive nematic in that it shows two static $+1/2$ defects facing head to head.
In contrast to the behavior of a passive system, however, in the ID state
the stationary defects' separation depends on the relative strength of activity and friction.
Note that, as the hydrodynamic screening through frictional forces increases and
the system further approaches the dry limit, the defects are pushed further apart and the ID state
occupies larger regions of the phase space. \par

\begin{figure}[!htb]
 \centering
 \includegraphics[width=0.93\columnwidth]{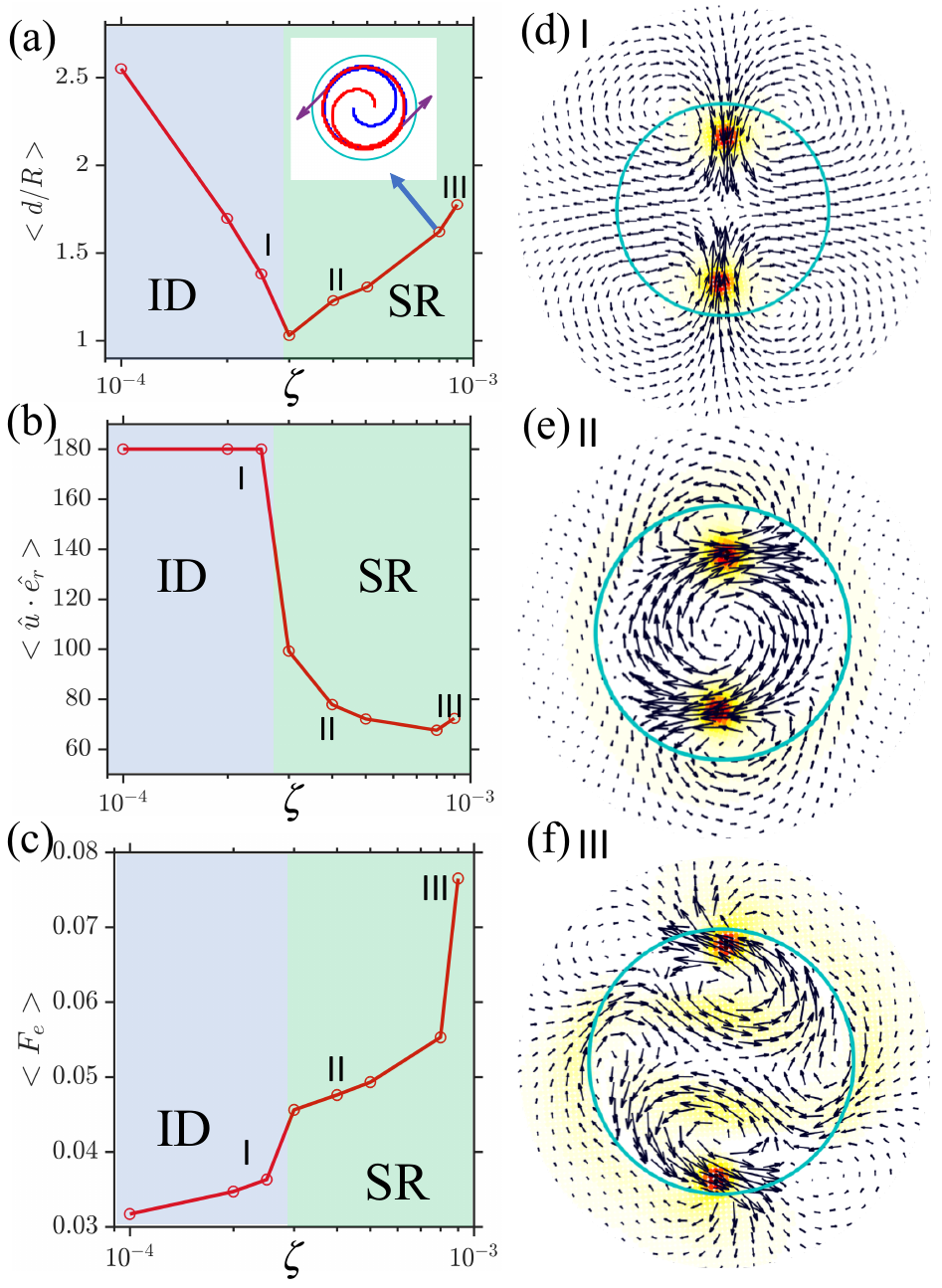}
 \caption{The effect of activity variation on (a) normalized defect separation, 
 (b) average angle of the defect orientation with the radial
  vector, and (c) average elastic free energy of the
  system, for ID and SR states. Points are sampled from the red line in Fig.~\ref{fig1}(b).
  (d-f) Velocity field in the domain corresponding to the
  points marked with \RomanNumeralCaps{1}-\RomanNumeralCaps{3}.
  The background color represents the nematic scalar order parameter ($S$);
  for the color scale refer to Fig.~\ref{fig1}(a).
  Sample trajectory of two steadily orbiting defects are shown in the inset of
  Fig.~\ref{fig2}(a) ($\zeta = 0.0008$, $f = 0.05$).}
 \label{fig2}
\end{figure}

Increasing activity leads to a decrease of the defect separation, as depicted in
Fig.~\ref{fig2}(a). 
The self-induced flows of the
defects for $\pm 1/2$ defects confined into the solid disk are illustrated in
Fig.~S5; one can appreciate
a polar double vortex structure for positive charge defects, which drives a
comet-like defect toward its head \cite{giomi2015}.
In the ID state the
director field is static. The active stresses, however, generate four equal-sized vortices of
deformed quadrupolar structure with a stagnation point in the centroid 
(Fig.~\ref{fig2}(d)).
Increasing activity pushes the defects toward each other, while
elastic, friction, and viscous forces keep them apart; all of these
forces cancel each other, leading to a static defects state.
Note that, despite the defects' lack of motion, active flows are strongest at the defects' core and,
consequently, viscous forces are imposed that oppose defect movement. \par

\begin{figure}[!htb]
 \centering
 \includegraphics[width=0.93\columnwidth]{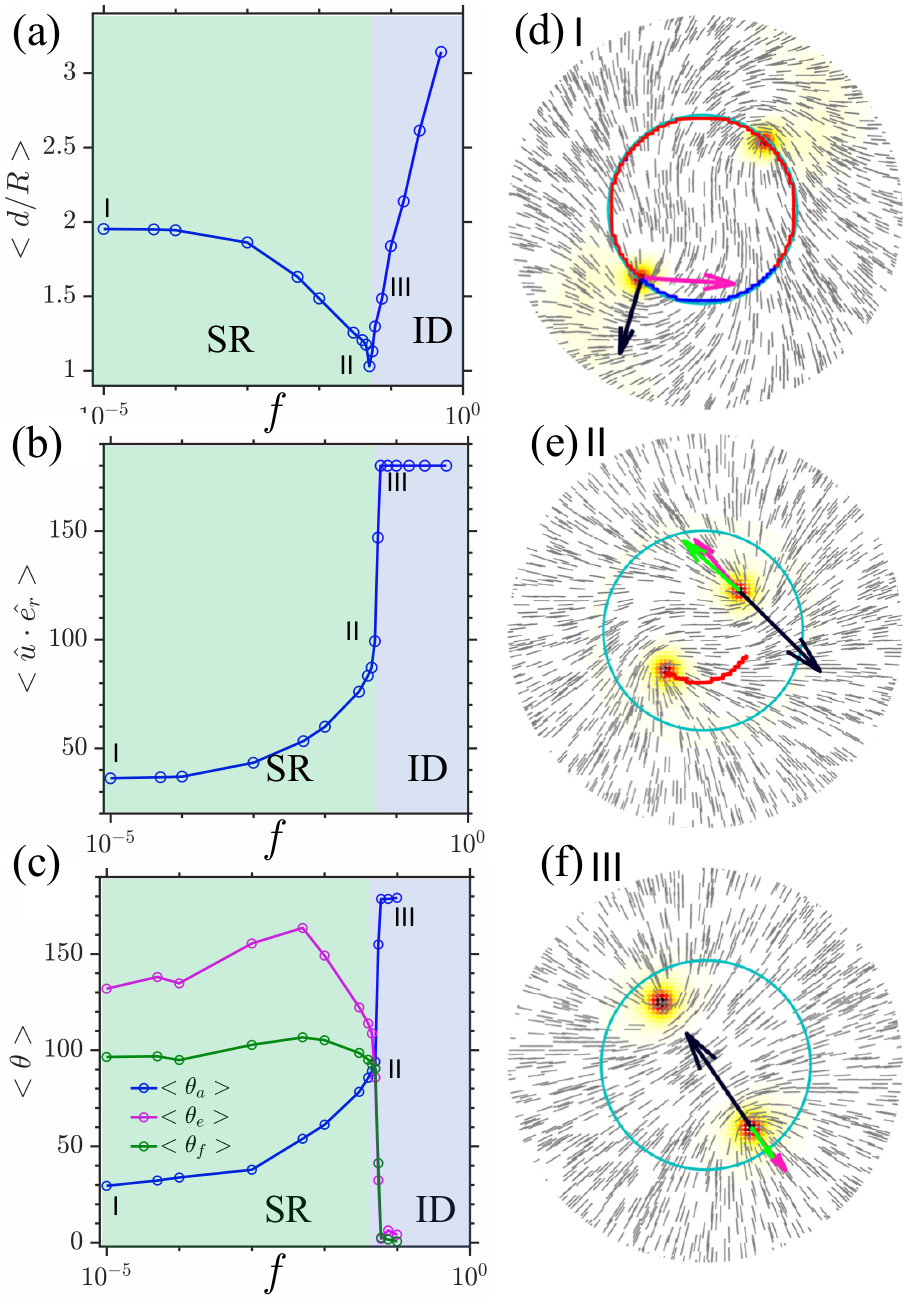}
 \caption{Effect of friction on ID and SR states. (a) Defect separation
  normalized by $R$, (b) average angle between the defect orientation and the radial
  vector, (c) average of the angles that the active, elastic, and friction forces exerted
  on the defects make with the radial vector; $\langle\theta_a\rangle$, $\langle\theta_e\rangle$, $\langle\theta_f\rangle$,
  respectively. Points are sampled from the blue line
  in Fig.~\ref{fig1}(b).
  (d-f) Director field and the trajectories of the defects corresponding to the
  points marked with \RomanNumeralCaps{1}-\RomanNumeralCaps{3}.
  Active, elastic, and frictional forces are denoted by black, pink, and green arrows, respectively.
  Panels (d),(e)
  correspond to the SR, where defects orbit in circles and, after the
  completion of each circle, retrace their steps. Only a part of the circular trajectories is plotted.
  Elastic and frictional forces in (e),(f) have been magnified by a factor of $2$.
  }
 \label{fig3}
\end{figure}

For a given value of friction, above a critical value of activity, the balance between the forces
on the defects is broken. Strong elastic
repulsions between $+1/2$ defects become dominant, and the active forces are unable to bring the
defects closer to each other. The defects tilt away from their line of centers (Fig.~\ref{fig2}(b)), and enter a
persistent circulation with random direction of orbiting, based on spontaneous symmetry-breaking 
(see the inset of Fig.~\ref{fig2}(a) for a representative trajectory of SR
and \href{run:../Movies/movie1.mp4}{Movie 1}). In contrast to the behavior of defects confined by hard and
impermeable walls, the transition to circulatory motion does not
require the creation of an additional pair of defects
\cite{norton2018}. The ease of reorientation of the director field at the soft activity boundary allows
the system to adapt
to the defects' director far-field and their underlying advective flows; circumventing the creation of a pair of new defects, giving
rise to a smooth transition from ID to SR.
Transitions from quiescence into a
moving state \cite{voituriez2005, woodhouse2012, ravnik2013, duclos2018}, and the emergence
of coherent circular motion \cite{segerer2015, liu2021}, are ubiquitous in active matter.
Here we have identified an additional lever for control by showing
that the right combination of activity level and friction can be used to tune the precise location
of the orbiting; it can be positioned, for example, right at the activity interface (Fig.~\ref{fig3}(d)).
Two of the four vortices in the ID state merge to form a larger vortex in the active region,
with two additional vortices farther away (Fig.~\ref{fig2}(e)).

In the SR state, increasing
activity not only drives the defects to orbits having a larger radius (by a stretching of the middle vortex 
to reduce the high shear)
but, due to the enhanced repulsion between them, they adopt an orientation along the radial
vector, trying to escape the active region.
In contrast, increasing the friction allows the
defects to rotate with a smaller orbiting radius, due to the suppression of momentum
propagation, the appearance of localized flows, and a reduction of the effective interaction between defects
(Fig.~\ref{fig3}(a) and Fig.~S6).

Figure~\ref{fig3} illustrates the role of friction for a constant value of the activity.
For low values of friction, where the defects' orbit is near the activity interface
(Fig.~\ref{fig3}(d) and \href{run:../Movies/movie2.mp4}{Movie 2}), the active forces are
closely oriented along the radial vector $\hat{e}_r$,
while the elastic forces form obtuse angles with the radial vector, keeping the
defects inside the active region. As the friction increases, these forces adopt a
perpendicular orientation with respect to the radial vector (Figs.~\ref{fig3}(c),(e) and
\href{run:../Movies/movie3.mp4}{Movie 3}) and, for
even larger values of friction, the defects become stationary, with forces parallel to the radial vector (Figs.~\ref{fig3} (c),(f)).
Interfacial active forces not only confine the defects inside
the active region, but also stabilize their sliding along the activity boundary. Defect steering
along the interface is also apparent for the case of a flat active/passive boundary
(Fig.~S2 and detailed discussions in the SI).   \par

In the SR state, increasing the activity reduces the nematic ordering, leading to an
increase of the elastic free energy
(Fig.~\ref{fig2}(c)). For sufficiently large values of friction and
activity (see transition from II to III in Fig.~\ref{fig2}, and \href{run:../Movies/movie4.mp4}{Movie 4}),
two stable elastic bands (walls) appear between two orbiting
defects, which move in phase with the defects. 
High frictional forces not only
allow the two bands that are formed to fit in the active
region, but also stabilize them, despite their high elastic cost.
Frictional forces reduce the shear stresses across the bands and, consequently, 
they can approach each other, reducing the effectiveness of
long-range hydrodynamic interactions \cite{thampi2014, doostmohammadi2016}. \par

\begin{figure*}[!htb]
 \centering
 \includegraphics[width=0.95\textwidth]{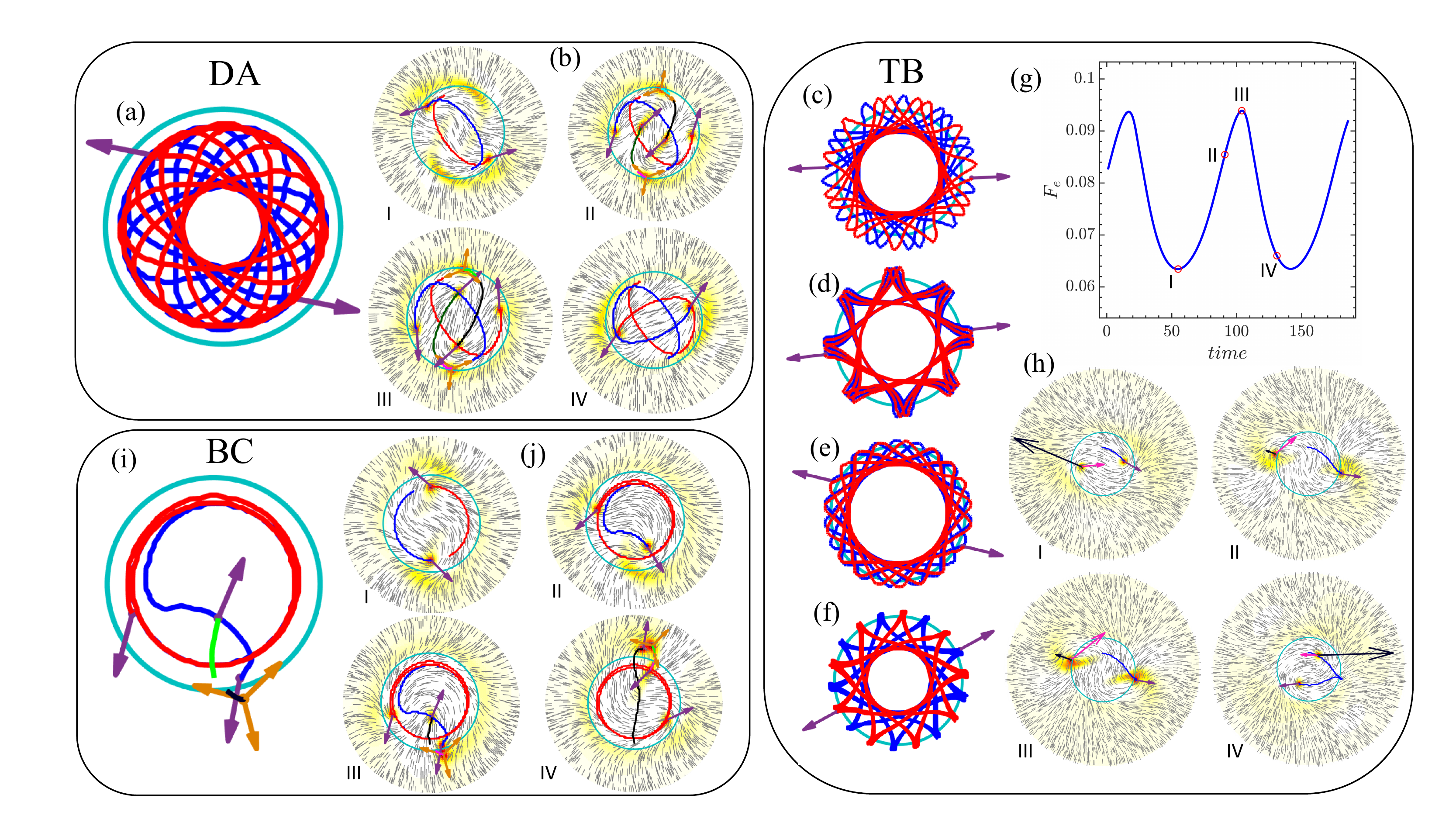}
 \caption{(a) Defect trajectory, and (b) snapshots of the director field in the dancing state 
 ($\zeta = 0.0016$, $f = 0.04$). 
  (c-f) Long-time defect trajectories
  for TB state for different values of activity ($\zeta = 0.0005, 0.0006, 0.0006, 0.0008$)
  and friction ($f = 0.0, 0.001, 0.002, 0.002$), respectively. Elastic free energy of
  the system (g) and snapshots of
  the director with forces exerted on the defect and the defect trajectory (blue)
  (h); (g)-(h) correspond to (f).
  Active, elastic, and frictional forces are shown with black, pink,
  and green arrows, respectively (frictional forces are relatively small).
  (i) Defect trajectories, and (j) snapshots of the
  director in the BC state ($\zeta = 0.001$, $f = 0.01$).
  The orientations of positive (negative) charge defects are identified by purple arrows
  (a triad of brown arrows). 
  Full blown animations of all of these states are provided
  in the SI \href{run:../Movies/movie5.mp4}{Movie 5}-\href{run:../Movies/movie9.mp4}{Movie 9}.}
 \label{fig4}
\end{figure*}

Along these bands, a Poiseuille-like flow is generated, leading to the
formation of additional vortices (Fig.~\ref{fig2}(f)).
If, in this state, the friction is reduced
or the activity is further increased, the elastic penalty incurred by the deformation of these bands
increases, causing the bands to unzip by creating a pair of defects on
each band (see elastic free energy amplification (reduction) during band formation
(unzipping) in \href{run:../Movies/movie5.mp4}{Movie 5}(c)).
While the newly liberated $+1/2$ defects move
along the bands to restore the nematic order, new $-1/2$ defects annihilate the already
existing $+1/2$ defects, and this cycle continues to form short-lived
dancing defects (\href{run:../Movies/movie5.mp4}{Movie 5}).
However, if the advective flows are large enough (further increasing the
activity),  band splitting occurs at the interface, where the stresses
are highest, and two fast-moving $+1/2$
defects from each band switch off and annihilate the newly created $-1/2$ ones.
The original $+1/2$ defects remain alive, forming a long-lived dancing state
(Figs.~\ref{fig4}(a),(b) and \href{run:../Movies/movie6.mp4}{Movie 6}). 

The TB region in the phase diagram occurs for intermediate values of activity and small
friction.
In this phase, the active
forces and the orientation of the defects form an acute angle, with radial vector
$\hat{e}_r$ (Figs.~\ref{fig3}(b),(c)), trying to
escape the active region. Once they reach the active/passive interface, the active forces drop
abruptly and induce a large unfavorable director distortion on the passive side.
The system's need to minimize elastic distortions creates elastic forces that
catapult the defects back to the active region, forming a ``yin yang'' structure
(Figs.~\ref{fig4}(g),(h) and \href{run:../Movies/movie8.mp4}{Movie 8}).
For higher values of activity, the elastic
distortion at the passive side next to the interface increases, and the catapulting effect is
stronger (faster) with a sharper reflection angle (Figs.~\ref{fig4}(e),(f) and 
\href{run:../Movies/movie7.mp4}{Movie 7}).
An inner, excluded circular region that defects never cross appears, due to
strong repulsive interactions. For higher values of activity, where the elastic-rebound
effect is sharper, this excluded area becomes smaller
(\href{run:../Movies/movie7.mp4}{Movie 7}). \par

Two contributions to the active forces, ${\bf{f}}^a = -(\zeta {\bm{\nabla}}\cdot {\bf{Q}} +
{\bm{\nabla}}\zeta \cdot {\bf{Q}})$,
arise from: i) the spatial variation of the director field which
is most appreciable in the proximity of topological defects and appears in the bulk of the
nematic (first term) and, ii) the large activity gradient ${\bm{\nabla}}\zeta$ at the activity interface (second term).
The second term, which distinguishes our work from that presented in past studies, is responsible for the
rich dynamical states presented here. The direction of this interfacial force depends on the angle between the
director field and the activity gradient; note the change of orientation of the force, from radially outward to inward when the
director field changes from radial to circular (Fig.~S3).
Also the
elastic reorienting torques that the defects exert on each other act to anti-align
\cite{vromans2016, shankar2018} them, further stabilizing the dynamics of the TB state. \par

A deeper understanding of the phase diagram can be gained through a theoretical analysis
that considers the force and torque balance when one $+1/2$ defect is near the flat
active/passive boundary (Fig.~S4).
By balancing
the elastic and active contributions to the torque and to the force normal to the boundary,
we find that a solution exists when the activity is below a threshold value, which corresponds
to the SR state in which defects can glide near the activity boundary. However, if the
activity reaches the threshold, a solution no longer exists, indicating that the defect cannot find
a steady state near the boundary - in correspondence to the TB state, where defects bounce
away as soon as they approach the interface.

We conclude our discussion with a brief mention of a peculiar behavior that we refer to as
BC, in which one of the defects undergoes a cruising behavior, and the other shows 
a bouncing behavior.
The viscous forces that drive the long range hydrodynamic interactions, and the frictional
forces, compete with each other; the resulting
screening length scale $\sqrt{\eta/f}$, defines the width of the elastic
bands that form within the active region.
Higher values of frictional damping are
required to observe the dancing state that is instigated by the pair of oppositely oriented
bands; the active region cannot accommodate two bands and, as such, the
system develops one band by breaking the symmetry of the flow and the director field.
Figures~\ref{fig4}(i),(j) show that while one of the defects shows a persistent circular motion,
the other one moves toward the interfacial region to induce a radially oriented
bending instability.
The newly formed band splits and creates a $-1/2$ defect, which rapidly annihilates the
original defect, while the new $+1/2$ defect is propelled away from the interface
(\href{run:../Movies/movie9.mp4}{Movie 9}).
This behavior shares some similarities with the experimental observation of
doubly periodic dynamics of defects, obtained from the slow nucleation
of defects at the boundary and fast circulating defects in
the bulk of microtubule-based active nematics under circular confinement
\cite{opathalage2019}. The creation of $-1/2$ defects at the solid wall might be indicative
of the slippery flow of microtubules at the disk surface, which induces tangential flows that, in turn,
drive bending instabilities and creation of a defect pair, similar to the BC state. \par

We have
described a new physical scenario in which the simple spatial patterning of
activity leads to defect confinement within a passive/active interface. When coupled to the control of friction at a substrate, we
have shown that it becomes possible to program defect motion,
thereby turning otherwise chaotic trajectories into precisely sculpted spatiotemporal states and
flow fields. \par

\begin{acknowledgments}
We thank Andrey Sokolov, Alexey Snezhko, 
Vincenzo Vitelli, and Xingzhou Tang for very helpful discussions.
This work on development of a fundamental understanding the structure and dynamics of active nematic 
liquid crystals is supported by the department of energy, basic energy sciences, division of materials science 
and engineering. The application of this work to the control of topological defects for logic operations is supported 
by the National Science Foundation, DMR-2011854.
R.Z. also acknowledges the financial
support of the Hong Kong RGC under grant no. 26302320. 
\end{acknowledgments}


\begin{thebibliography}{68}%
\makeatletter
\providecommand \@ifxundefined [1]{%
 \@ifx{#1\undefined}
}%
\providecommand \@ifnum [1]{%
 \ifnum #1\expandafter \@firstoftwo
 \else \expandafter \@secondoftwo
 \fi
}%
\providecommand \@ifx [1]{%
 \ifx #1\expandafter \@firstoftwo
 \else \expandafter \@secondoftwo
 \fi
}%
\providecommand \natexlab [1]{#1}%
\providecommand \enquote  [1]{``#1''}%
\providecommand \bibnamefont  [1]{#1}%
\providecommand \bibfnamefont [1]{#1}%
\providecommand \citenamefont [1]{#1}%
\providecommand \href@noop [0]{\@secondoftwo}%
\providecommand \href [0]{\begingroup \@sanitize@url \@href}%
\providecommand \@href[1]{\@@startlink{#1}\@@href}%
\providecommand \@@href[1]{\endgroup#1\@@endlink}%
\providecommand \@sanitize@url [0]{\catcode `\\12\catcode `\$12\catcode
  `\&12\catcode `\#12\catcode `\^12\catcode `\_12\catcode `\%12\relax}%
\providecommand \@@startlink[1]{}%
\providecommand \@@endlink[0]{}%
\providecommand \url  [0]{\begingroup\@sanitize@url \@url }%
\providecommand \@url [1]{\endgroup\@href {#1}{\urlprefix }}%
\providecommand \urlprefix  [0]{URL }%
\providecommand \Eprint [0]{\href }%
\providecommand \doibase [0]{http://dx.doi.org/}%
\providecommand \selectlanguage [0]{\@gobble}%
\providecommand \bibinfo  [0]{\@secondoftwo}%
\providecommand \bibfield  [0]{\@secondoftwo}%
\providecommand \translation [1]{[#1]}%
\providecommand \BibitemOpen [0]{}%
\providecommand \bibitemStop [0]{}%
\providecommand \bibitemNoStop [0]{.\EOS\space}%
\providecommand \EOS [0]{\spacefactor3000\relax}%
\providecommand \BibitemShut  [1]{\csname bibitem#1\endcsname}%
\let\auto@bib@innerbib\@empty
\bibitem [{\citenamefont {Ramaswamy}(2010)}]{ramaswamy2010}%
  \BibitemOpen
  \bibfield  {author} {\bibinfo {author} {\bibfnamefont {S.}~\bibnamefont
  {Ramaswamy}},\ }\href {\doibase
  https://doi.org/10.1146/annurev-conmatphys-070909-104101} {\bibfield
  {journal} {\bibinfo  {journal} {Annu. Rev. Condens. Matter Phys.}\ }\textbf
  {\bibinfo {volume} {1}},\ \bibinfo {pages} {323} (\bibinfo {year}
  {2010})}\BibitemShut {NoStop}%
\bibitem [{\citenamefont {Marchetti}\ \emph {et~al.}(2013)\citenamefont
  {Marchetti}, \citenamefont {Joanny}, \citenamefont {Ramaswamy}, \citenamefont
  {Liverpool}, \citenamefont {Prost}, \citenamefont {Rao},\ and\ \citenamefont
  {Simha}}]{marchetti2013}%
  \BibitemOpen
  \bibfield  {author} {\bibinfo {author} {\bibfnamefont {M.~C.}\ \bibnamefont
  {Marchetti}}, \bibinfo {author} {\bibfnamefont {J.-F.}\ \bibnamefont
  {Joanny}}, \bibinfo {author} {\bibfnamefont {S.}~\bibnamefont {Ramaswamy}},
  \bibinfo {author} {\bibfnamefont {T.~B.}\ \bibnamefont {Liverpool}}, \bibinfo
  {author} {\bibfnamefont {J.}~\bibnamefont {Prost}}, \bibinfo {author}
  {\bibfnamefont {M.}~\bibnamefont {Rao}}, \ and\ \bibinfo {author}
  {\bibfnamefont {R.~A.}\ \bibnamefont {Simha}},\ }\href {\doibase
  10.1103/RevModPhys.85.1143} {\bibfield  {journal} {\bibinfo  {journal} {Rev.
  Mod. Phys.}\ }\textbf {\bibinfo {volume} {85}},\ \bibinfo {pages} {1143}
  (\bibinfo {year} {2013})}\BibitemShut {NoStop}%
\bibitem [{\citenamefont {Needleman}\ and\ \citenamefont
  {Dogic}(2017)}]{needleman2017}%
  \BibitemOpen
  \bibfield  {author} {\bibinfo {author} {\bibfnamefont {D.}~\bibnamefont
  {Needleman}}\ and\ \bibinfo {author} {\bibfnamefont {Z.}~\bibnamefont
  {Dogic}},\ }\href {\doibase https://doi.org/10.1038/natrevmats.2017.48}
  {\bibfield  {journal} {\bibinfo  {journal} {Nat. Rev. Mater.}\ }\textbf
  {\bibinfo {volume} {2}},\ \bibinfo {pages} {17048} (\bibinfo {year}
  {2017})}\BibitemShut {NoStop}%
\bibitem [{\citenamefont {Doostmohammadi}\ \emph {et~al.}(2018)\citenamefont
  {Doostmohammadi}, \citenamefont {Ign{\'e}s-Mullol}, \citenamefont {Yeomans},\
  and\ \citenamefont {Sagu{\'e}s}}]{doostmohammadi2018}%
  \BibitemOpen
  \bibfield  {author} {\bibinfo {author} {\bibfnamefont {A.}~\bibnamefont
  {Doostmohammadi}}, \bibinfo {author} {\bibfnamefont {J.}~\bibnamefont
  {Ign{\'e}s-Mullol}}, \bibinfo {author} {\bibfnamefont {J.~M.}\ \bibnamefont
  {Yeomans}}, \ and\ \bibinfo {author} {\bibfnamefont {F.}~\bibnamefont
  {Sagu{\'e}s}},\ }\href {\doibase https://doi.org/10.1038/s41467-018-05666-8}
  {\bibfield  {journal} {\bibinfo  {journal} {Nat. Commun.}\ }\textbf {\bibinfo
  {volume} {9}},\ \bibinfo {pages} {3246} (\bibinfo {year} {2018})}\BibitemShut
  {NoStop}%
\bibitem [{\citenamefont {Shaebani}\ \emph {et~al.}(2020)\citenamefont
  {Shaebani}, \citenamefont {Wysocki}, \citenamefont {Winkler}, \citenamefont
  {Gompper},\ and\ \citenamefont {Rieger}}]{shaebani2020}%
  \BibitemOpen
  \bibfield  {author} {\bibinfo {author} {\bibfnamefont {M.~R.}\ \bibnamefont
  {Shaebani}}, \bibinfo {author} {\bibfnamefont {A.}~\bibnamefont {Wysocki}},
  \bibinfo {author} {\bibfnamefont {R.~G.}\ \bibnamefont {Winkler}}, \bibinfo
  {author} {\bibfnamefont {G.}~\bibnamefont {Gompper}}, \ and\ \bibinfo
  {author} {\bibfnamefont {H.}~\bibnamefont {Rieger}},\ }\href {\doibase
  https://doi.org/10.1038/s42254-020-0152-1} {\bibfield  {journal} {\bibinfo
  {journal} {Nat. Rev. Phys.}\ }\textbf {\bibinfo {volume} {2}},\ \bibinfo
  {pages} {181} (\bibinfo {year} {2020})}\BibitemShut {NoStop}%
\bibitem [{\citenamefont {Ign{\'e}s-Mullol}\ and\ \citenamefont
  {Sagu{\'e}s}(2020)}]{ignes2020}%
  \BibitemOpen
  \bibfield  {author} {\bibinfo {author} {\bibfnamefont {J.}~\bibnamefont
  {Ign{\'e}s-Mullol}}\ and\ \bibinfo {author} {\bibfnamefont {F.}~\bibnamefont
  {Sagu{\'e}s}},\ }\href {\doibase https://doi.org/10.1016/j.cocis.2020.04.007}
  {\bibfield  {journal} {\bibinfo  {journal} {Curr. Opin. Colloid Interface
  Sci.}\ }\textbf {\bibinfo {volume} {49}},\ \bibinfo {pages} {16} (\bibinfo
  {year} {2020})}\BibitemShut {NoStop}%
\bibitem [{\citenamefont {Zhang}\ \emph
  {et~al.}(2021{\natexlab{a}})\citenamefont {Zhang}, \citenamefont
  {Mozaffari},\ and\ \citenamefont {de~Pablo}}]{zhang2021auto}%
  \BibitemOpen
  \bibfield  {author} {\bibinfo {author} {\bibfnamefont {R.}~\bibnamefont
  {Zhang}}, \bibinfo {author} {\bibfnamefont {A.}~\bibnamefont {Mozaffari}}, \
  and\ \bibinfo {author} {\bibfnamefont {J.~J.}\ \bibnamefont {de~Pablo}},\
  }\href {\doibase https://doi.org/10.1038/s41578-020-00272-x} {\bibfield
  {journal} {\bibinfo  {journal} {Nat. Rev. Mater.}\ }\textbf {\bibinfo
  {volume} {6}},\ \bibinfo {pages} {437} (\bibinfo {year}
  {2021}{\natexlab{a}})}\BibitemShut {NoStop}%
\bibitem [{\citenamefont {Sanchez}\ \emph {et~al.}(2012)\citenamefont
  {Sanchez}, \citenamefont {Chen}, \citenamefont {DeCamp}, \citenamefont
  {Heymann},\ and\ \citenamefont {Dogic}}]{sanchez2012}%
  \BibitemOpen
  \bibfield  {author} {\bibinfo {author} {\bibfnamefont {T.}~\bibnamefont
  {Sanchez}}, \bibinfo {author} {\bibfnamefont {D.~T.}\ \bibnamefont {Chen}},
  \bibinfo {author} {\bibfnamefont {S.~J.}\ \bibnamefont {DeCamp}}, \bibinfo
  {author} {\bibfnamefont {M.}~\bibnamefont {Heymann}}, \ and\ \bibinfo
  {author} {\bibfnamefont {Z.}~\bibnamefont {Dogic}},\ }\href {\doibase
  10.1038/nature11591} {\bibfield  {journal} {\bibinfo  {journal} {Nature}\
  }\textbf {\bibinfo {volume} {491}},\ \bibinfo {pages} {431} (\bibinfo {year}
  {2012})}\BibitemShut {NoStop}%
\bibitem [{\citenamefont {Keber}\ \emph {et~al.}(2014)\citenamefont {Keber},
  \citenamefont {Loiseau}, \citenamefont {Sanchez}, \citenamefont {DeCamp},
  \citenamefont {Giomi}, \citenamefont {Bowick}, \citenamefont {Marchetti},
  \citenamefont {Dogic},\ and\ \citenamefont {Bausch}}]{keber2014}%
  \BibitemOpen
  \bibfield  {author} {\bibinfo {author} {\bibfnamefont {F.~C.}\ \bibnamefont
  {Keber}}, \bibinfo {author} {\bibfnamefont {E.}~\bibnamefont {Loiseau}},
  \bibinfo {author} {\bibfnamefont {T.}~\bibnamefont {Sanchez}}, \bibinfo
  {author} {\bibfnamefont {S.~J.}\ \bibnamefont {DeCamp}}, \bibinfo {author}
  {\bibfnamefont {L.}~\bibnamefont {Giomi}}, \bibinfo {author} {\bibfnamefont
  {M.~J.}\ \bibnamefont {Bowick}}, \bibinfo {author} {\bibfnamefont {M.~C.}\
  \bibnamefont {Marchetti}}, \bibinfo {author} {\bibfnamefont {Z.}~\bibnamefont
  {Dogic}}, \ and\ \bibinfo {author} {\bibfnamefont {A.~R.}\ \bibnamefont
  {Bausch}},\ }\href {\doibase https://doi.org/10.1126/science.1254784}
  {\bibfield  {journal} {\bibinfo  {journal} {Science}\ }\textbf {\bibinfo
  {volume} {345}},\ \bibinfo {pages} {1135} (\bibinfo {year}
  {2014})}\BibitemShut {NoStop}%
\bibitem [{\citenamefont {Kumar}\ \emph {et~al.}(2018)\citenamefont {Kumar},
  \citenamefont {Zhang}, \citenamefont {de~Pablo},\ and\ \citenamefont
  {Gardel}}]{kumar2018}%
  \BibitemOpen
  \bibfield  {author} {\bibinfo {author} {\bibfnamefont {N.}~\bibnamefont
  {Kumar}}, \bibinfo {author} {\bibfnamefont {R.}~\bibnamefont {Zhang}},
  \bibinfo {author} {\bibfnamefont {J.~J.}\ \bibnamefont {de~Pablo}}, \ and\
  \bibinfo {author} {\bibfnamefont {M.~L.}\ \bibnamefont {Gardel}},\ }\href
  {\doibase 10.1126/sciadv.aat7779} {\bibfield  {journal} {\bibinfo  {journal}
  {Sci. Adv.}\ }\textbf {\bibinfo {volume} {4}},\ \bibinfo {pages} {eaat7779}
  (\bibinfo {year} {2018})}\BibitemShut {NoStop}%
\bibitem [{\citenamefont {Rivas}\ \emph {et~al.}(2020)\citenamefont {Rivas},
  \citenamefont {Shendruk}, \citenamefont {Henry}, \citenamefont {Reich},\ and\
  \citenamefont {Leheny}}]{rivas2020}%
  \BibitemOpen
  \bibfield  {author} {\bibinfo {author} {\bibfnamefont {D.~P.}\ \bibnamefont
  {Rivas}}, \bibinfo {author} {\bibfnamefont {T.~N.}\ \bibnamefont {Shendruk}},
  \bibinfo {author} {\bibfnamefont {R.~R.}\ \bibnamefont {Henry}}, \bibinfo
  {author} {\bibfnamefont {D.~H.}\ \bibnamefont {Reich}}, \ and\ \bibinfo
  {author} {\bibfnamefont {R.~L.}\ \bibnamefont {Leheny}},\ }\href {\doibase
  10.1039/D0SM00693A} {\bibfield  {journal} {\bibinfo  {journal} {Soft Matter}\
  }\textbf {\bibinfo {volume} {16}},\ \bibinfo {pages} {9331} (\bibinfo {year}
  {2020})}\BibitemShut {NoStop}%
\bibitem [{\citenamefont {Dunkel}\ \emph {et~al.}(2013)\citenamefont {Dunkel},
  \citenamefont {Heidenreich}, \citenamefont {Drescher}, \citenamefont
  {Wensink}, \citenamefont {B{\"a}r},\ and\ \citenamefont
  {Goldstein}}]{dunkel2013}%
  \BibitemOpen
  \bibfield  {author} {\bibinfo {author} {\bibfnamefont {J.}~\bibnamefont
  {Dunkel}}, \bibinfo {author} {\bibfnamefont {S.}~\bibnamefont {Heidenreich}},
  \bibinfo {author} {\bibfnamefont {K.}~\bibnamefont {Drescher}}, \bibinfo
  {author} {\bibfnamefont {H.~H.}\ \bibnamefont {Wensink}}, \bibinfo {author}
  {\bibfnamefont {M.}~\bibnamefont {B{\"a}r}}, \ and\ \bibinfo {author}
  {\bibfnamefont {R.~E.}\ \bibnamefont {Goldstein}},\ }\href {\doibase
  10.1103/PhysRevLett.110.228102} {\bibfield  {journal} {\bibinfo  {journal}
  {Phys. Rev. Lett.}\ }\textbf {\bibinfo {volume} {110}},\ \bibinfo {pages}
  {228102} (\bibinfo {year} {2013})}\BibitemShut {NoStop}%
\bibitem [{\citenamefont {Li}\ \emph {et~al.}(2019)\citenamefont {Li},
  \citenamefont {Shi}, \citenamefont {Huang}, \citenamefont {Chen},
  \citenamefont {Xiao}, \citenamefont {Liu}, \citenamefont {Chat{\'e}},\ and\
  \citenamefont {Zhang}}]{li2019}%
  \BibitemOpen
  \bibfield  {author} {\bibinfo {author} {\bibfnamefont {H.}~\bibnamefont
  {Li}}, \bibinfo {author} {\bibfnamefont {X.-q.}\ \bibnamefont {Shi}},
  \bibinfo {author} {\bibfnamefont {M.}~\bibnamefont {Huang}}, \bibinfo
  {author} {\bibfnamefont {X.}~\bibnamefont {Chen}}, \bibinfo {author}
  {\bibfnamefont {M.}~\bibnamefont {Xiao}}, \bibinfo {author} {\bibfnamefont
  {C.}~\bibnamefont {Liu}}, \bibinfo {author} {\bibfnamefont {H.}~\bibnamefont
  {Chat{\'e}}}, \ and\ \bibinfo {author} {\bibfnamefont {H.}~\bibnamefont
  {Zhang}},\ }\href {\doibase https://doi.org/10.1073/pnas.1812570116}
  {\bibfield  {journal} {\bibinfo  {journal} {Proc. Natl. Acad. Sci. U.S.A.}\
  }\textbf {\bibinfo {volume} {116}},\ \bibinfo {pages} {777} (\bibinfo {year}
  {2019})}\BibitemShut {NoStop}%
\bibitem [{\citenamefont {Zhou}\ \emph {et~al.}(2014)\citenamefont {Zhou},
  \citenamefont {Sokolov}, \citenamefont {Lavrentovich},\ and\ \citenamefont
  {Aranson}}]{zhou2014}%
  \BibitemOpen
  \bibfield  {author} {\bibinfo {author} {\bibfnamefont {S.}~\bibnamefont
  {Zhou}}, \bibinfo {author} {\bibfnamefont {A.}~\bibnamefont {Sokolov}},
  \bibinfo {author} {\bibfnamefont {O.~D.}\ \bibnamefont {Lavrentovich}}, \
  and\ \bibinfo {author} {\bibfnamefont {I.~S.}\ \bibnamefont {Aranson}},\
  }\href {\doibase https://doi.org/10.1073/pnas.1321926111} {\bibfield
  {journal} {\bibinfo  {journal} {Proc. Natl. Acad. Sci. U.S.A.}\ }\textbf
  {\bibinfo {volume} {111}},\ \bibinfo {pages} {1265} (\bibinfo {year}
  {2014})}\BibitemShut {NoStop}%
\bibitem [{\citenamefont {Duclos}\ \emph {et~al.}(2017)\citenamefont {Duclos},
  \citenamefont {Erlenk{\"a}mper}, \citenamefont {Joanny},\ and\ \citenamefont
  {Silberzan}}]{duclos2017}%
  \BibitemOpen
  \bibfield  {author} {\bibinfo {author} {\bibfnamefont {G.}~\bibnamefont
  {Duclos}}, \bibinfo {author} {\bibfnamefont {C.}~\bibnamefont
  {Erlenk{\"a}mper}}, \bibinfo {author} {\bibfnamefont {J.-F.}\ \bibnamefont
  {Joanny}}, \ and\ \bibinfo {author} {\bibfnamefont {P.}~\bibnamefont
  {Silberzan}},\ }\href {\doibase https://doi.org/10.1038/nphys3876} {\bibfield
   {journal} {\bibinfo  {journal} {Nat. Phys.}\ }\textbf {\bibinfo {volume}
  {13}},\ \bibinfo {pages} {58} (\bibinfo {year} {2017})}\BibitemShut {NoStop}%
\bibitem [{\citenamefont {Kawaguchi}\ \emph {et~al.}(2017)\citenamefont
  {Kawaguchi}, \citenamefont {Kageyama},\ and\ \citenamefont
  {Sano}}]{kawaguchi2017}%
  \BibitemOpen
  \bibfield  {author} {\bibinfo {author} {\bibfnamefont {K.}~\bibnamefont
  {Kawaguchi}}, \bibinfo {author} {\bibfnamefont {R.}~\bibnamefont {Kageyama}},
  \ and\ \bibinfo {author} {\bibfnamefont {M.}~\bibnamefont {Sano}},\ }\href
  {\doibase https://doi.org/10.1038/nature22321} {\bibfield  {journal}
  {\bibinfo  {journal} {Nature}\ }\textbf {\bibinfo {volume} {545}},\ \bibinfo
  {pages} {327} (\bibinfo {year} {2017})}\BibitemShut {NoStop}%
\bibitem [{\citenamefont {Blanch-Mercader}\ \emph {et~al.}(2018)\citenamefont
  {Blanch-Mercader}, \citenamefont {Yashunsky}, \citenamefont {Garcia},
  \citenamefont {Duclos}, \citenamefont {Giomi},\ and\ \citenamefont
  {Silberzan}}]{blanch2018}%
  \BibitemOpen
  \bibfield  {author} {\bibinfo {author} {\bibfnamefont {C.}~\bibnamefont
  {Blanch-Mercader}}, \bibinfo {author} {\bibfnamefont {V.}~\bibnamefont
  {Yashunsky}}, \bibinfo {author} {\bibfnamefont {S.}~\bibnamefont {Garcia}},
  \bibinfo {author} {\bibfnamefont {G.}~\bibnamefont {Duclos}}, \bibinfo
  {author} {\bibfnamefont {L.}~\bibnamefont {Giomi}}, \ and\ \bibinfo {author}
  {\bibfnamefont {P.}~\bibnamefont {Silberzan}},\ }\href {\doibase
  10.1103/PhysRevLett.120.208101} {\bibfield  {journal} {\bibinfo  {journal}
  {Phys. Rev. Lett.}\ }\textbf {\bibinfo {volume} {120}},\ \bibinfo {pages}
  {208101} (\bibinfo {year} {2018})}\BibitemShut {NoStop}%
\bibitem [{\citenamefont {Dell'Arciprete}\ \emph {et~al.}(2018)\citenamefont
  {Dell'Arciprete}, \citenamefont {Blow}, \citenamefont {Brown}, \citenamefont
  {Farrell}, \citenamefont {Lintuvuori}, \citenamefont {McVey}, \citenamefont
  {Marenduzzo},\ and\ \citenamefont {Poon}}]{dell2018}%
  \BibitemOpen
  \bibfield  {author} {\bibinfo {author} {\bibfnamefont {D.}~\bibnamefont
  {Dell'Arciprete}}, \bibinfo {author} {\bibfnamefont {M.}~\bibnamefont
  {Blow}}, \bibinfo {author} {\bibfnamefont {A.}~\bibnamefont {Brown}},
  \bibinfo {author} {\bibfnamefont {F.}~\bibnamefont {Farrell}}, \bibinfo
  {author} {\bibfnamefont {J.~S.}\ \bibnamefont {Lintuvuori}}, \bibinfo
  {author} {\bibfnamefont {A.}~\bibnamefont {McVey}}, \bibinfo {author}
  {\bibfnamefont {D.}~\bibnamefont {Marenduzzo}}, \ and\ \bibinfo {author}
  {\bibfnamefont {W.~C.}\ \bibnamefont {Poon}},\ }\href {\doibase
  https://doi.org/10.1038/s41467-018-06370-3} {\bibfield  {journal} {\bibinfo
  {journal} {Nat. Commun.}\ }\textbf {\bibinfo {volume} {9}},\ \bibinfo {pages}
  {4190} (\bibinfo {year} {2018})}\BibitemShut {NoStop}%
\bibitem [{\citenamefont {Simha}\ and\ \citenamefont
  {Ramaswamy}(2002)}]{simha2002}%
  \BibitemOpen
  \bibfield  {author} {\bibinfo {author} {\bibfnamefont {R.~A.}\ \bibnamefont
  {Simha}}\ and\ \bibinfo {author} {\bibfnamefont {S.}~\bibnamefont
  {Ramaswamy}},\ }\href {\doibase 10.1103/PhysRevLett.89.058101} {\bibfield
  {journal} {\bibinfo  {journal} {Phys. Rev. Lett.}\ }\textbf {\bibinfo
  {volume} {89}},\ \bibinfo {pages} {058101} (\bibinfo {year}
  {2002})}\BibitemShut {NoStop}%
\bibitem [{\citenamefont {Narayan}\ \emph {et~al.}(2007)\citenamefont
  {Narayan}, \citenamefont {Ramaswamy},\ and\ \citenamefont
  {Menon}}]{narayan2007}%
  \BibitemOpen
  \bibfield  {author} {\bibinfo {author} {\bibfnamefont {V.}~\bibnamefont
  {Narayan}}, \bibinfo {author} {\bibfnamefont {S.}~\bibnamefont {Ramaswamy}},
  \ and\ \bibinfo {author} {\bibfnamefont {N.}~\bibnamefont {Menon}},\ }\href
  {\doibase https://doi.org/10.1126/science.1140414} {\bibfield  {journal}
  {\bibinfo  {journal} {Science}\ }\textbf {\bibinfo {volume} {317}},\ \bibinfo
  {pages} {105} (\bibinfo {year} {2007})}\BibitemShut {NoStop}%
\bibitem [{\citenamefont {Marenduzzo}\ \emph {et~al.}(2007)\citenamefont
  {Marenduzzo}, \citenamefont {Orlandini},\ and\ \citenamefont
  {Yeomans}}]{marenduzzo2007}%
  \BibitemOpen
  \bibfield  {author} {\bibinfo {author} {\bibfnamefont {D.}~\bibnamefont
  {Marenduzzo}}, \bibinfo {author} {\bibfnamefont {E.}~\bibnamefont
  {Orlandini}}, \ and\ \bibinfo {author} {\bibfnamefont {J.~M.}\ \bibnamefont
  {Yeomans}},\ }\href {\doibase 10.1103/PhysRevLett.98.118102} {\bibfield
  {journal} {\bibinfo  {journal} {Phys. Rev. Lett.}\ }\textbf {\bibinfo
  {volume} {98}},\ \bibinfo {pages} {118102} (\bibinfo {year}
  {2007})}\BibitemShut {NoStop}%
\bibitem [{\citenamefont {Giomi}\ \emph {et~al.}(2011)\citenamefont {Giomi},
  \citenamefont {Mahadevan}, \citenamefont {Chakraborty},\ and\ \citenamefont
  {Hagan}}]{giomi2011}%
  \BibitemOpen
  \bibfield  {author} {\bibinfo {author} {\bibfnamefont {L.}~\bibnamefont
  {Giomi}}, \bibinfo {author} {\bibfnamefont {L.}~\bibnamefont {Mahadevan}},
  \bibinfo {author} {\bibfnamefont {B.}~\bibnamefont {Chakraborty}}, \ and\
  \bibinfo {author} {\bibfnamefont {M.~F.}\ \bibnamefont {Hagan}},\ }\href
  {\doibase 10.1103/PhysRevLett.106.218101} {\bibfield  {journal} {\bibinfo
  {journal} {Phys. Rev. Lett.}\ }\textbf {\bibinfo {volume} {106}},\ \bibinfo
  {pages} {218101} (\bibinfo {year} {2011})}\BibitemShut {NoStop}%
\bibitem [{\citenamefont {Giomi}\ \emph {et~al.}(2013)\citenamefont {Giomi},
  \citenamefont {Bowick}, \citenamefont {Ma},\ and\ \citenamefont
  {Marchetti}}]{giomi2013}%
  \BibitemOpen
  \bibfield  {author} {\bibinfo {author} {\bibfnamefont {L.}~\bibnamefont
  {Giomi}}, \bibinfo {author} {\bibfnamefont {M.~J.}\ \bibnamefont {Bowick}},
  \bibinfo {author} {\bibfnamefont {X.}~\bibnamefont {Ma}}, \ and\ \bibinfo
  {author} {\bibfnamefont {M.~C.}\ \bibnamefont {Marchetti}},\ }\href {\doibase
  10.1103/PhysRevLett.110.228101} {\bibfield  {journal} {\bibinfo  {journal}
  {Phys. Rev. Lett.}\ }\textbf {\bibinfo {volume} {110}},\ \bibinfo {pages}
  {228101} (\bibinfo {year} {2013})}\BibitemShut {NoStop}%
\bibitem [{\citenamefont {Giomi}(2015)}]{giomi2015}%
  \BibitemOpen
  \bibfield  {author} {\bibinfo {author} {\bibfnamefont {L.}~\bibnamefont
  {Giomi}},\ }\href {\doibase https://doi.org/10.1103/PhysRevX.5.031003}
  {\bibfield  {journal} {\bibinfo  {journal} {Phys. Rev. X}\ }\textbf {\bibinfo
  {volume} {5}},\ \bibinfo {pages} {031003} (\bibinfo {year}
  {2015})}\BibitemShut {NoStop}%
\bibitem [{\citenamefont {Hemingway}\ \emph {et~al.}(2016)\citenamefont
  {Hemingway}, \citenamefont {Mishra}, \citenamefont {Marchetti},\ and\
  \citenamefont {Fielding}}]{hemingway2016}%
  \BibitemOpen
  \bibfield  {author} {\bibinfo {author} {\bibfnamefont {E.~J.}\ \bibnamefont
  {Hemingway}}, \bibinfo {author} {\bibfnamefont {P.}~\bibnamefont {Mishra}},
  \bibinfo {author} {\bibfnamefont {M.~C.}\ \bibnamefont {Marchetti}}, \ and\
  \bibinfo {author} {\bibfnamefont {S.~M.}\ \bibnamefont {Fielding}},\ }\href
  {\doibase https://doi.org/10.1039/C6SM00812G} {\bibfield  {journal} {\bibinfo
   {journal} {Soft Matter}\ }\textbf {\bibinfo {volume} {12}},\ \bibinfo
  {pages} {7943} (\bibinfo {year} {2016})}\BibitemShut {NoStop}%
\bibitem [{\citenamefont {Doostmohammadi}\ \emph
  {et~al.}(2016{\natexlab{a}})\citenamefont {Doostmohammadi}, \citenamefont
  {Thampi},\ and\ \citenamefont {Yeomans}}]{doostmohammadi2016b}%
  \BibitemOpen
  \bibfield  {author} {\bibinfo {author} {\bibfnamefont {A.}~\bibnamefont
  {Doostmohammadi}}, \bibinfo {author} {\bibfnamefont {S.~P.}\ \bibnamefont
  {Thampi}}, \ and\ \bibinfo {author} {\bibfnamefont {J.~M.}\ \bibnamefont
  {Yeomans}},\ }\href {\doibase https://doi.org/10.1103/PhysRevLett.117.048102}
  {\bibfield  {journal} {\bibinfo  {journal} {Phys. Rev. Lett.}\ }\textbf
  {\bibinfo {volume} {117}},\ \bibinfo {pages} {048102} (\bibinfo {year}
  {2016}{\natexlab{a}})}\BibitemShut {NoStop}%
\bibitem [{\citenamefont {Shankar}\ \emph {et~al.}(2018)\citenamefont
  {Shankar}, \citenamefont {Ramaswamy}, \citenamefont {Marchetti},\ and\
  \citenamefont {Bowick}}]{shankar2018}%
  \BibitemOpen
  \bibfield  {author} {\bibinfo {author} {\bibfnamefont {S.}~\bibnamefont
  {Shankar}}, \bibinfo {author} {\bibfnamefont {S.}~\bibnamefont {Ramaswamy}},
  \bibinfo {author} {\bibfnamefont {M.~C.}\ \bibnamefont {Marchetti}}, \ and\
  \bibinfo {author} {\bibfnamefont {M.~J.}\ \bibnamefont {Bowick}},\ }\href
  {\doibase 10.1103/PhysRevLett.121.108002} {\bibfield  {journal} {\bibinfo
  {journal} {Phys. Rev. Lett.}\ }\textbf {\bibinfo {volume} {121}},\ \bibinfo
  {pages} {108002} (\bibinfo {year} {2018})}\BibitemShut {NoStop}%
\bibitem [{\citenamefont {Norton}\ \emph {et~al.}(2020)\citenamefont {Norton},
  \citenamefont {Grover}, \citenamefont {Hagan},\ and\ \citenamefont
  {Fraden}}]{norton2020}%
  \BibitemOpen
  \bibfield  {author} {\bibinfo {author} {\bibfnamefont {M.~M.}\ \bibnamefont
  {Norton}}, \bibinfo {author} {\bibfnamefont {P.}~\bibnamefont {Grover}},
  \bibinfo {author} {\bibfnamefont {M.~F.}\ \bibnamefont {Hagan}}, \ and\
  \bibinfo {author} {\bibfnamefont {S.}~\bibnamefont {Fraden}},\ }\href
  {\doibase https://doi.org/10.1103/PhysRevLett.125.178005} {\bibfield
  {journal} {\bibinfo  {journal} {Phys. Rev. Lett.}\ }\textbf {\bibinfo
  {volume} {125}},\ \bibinfo {pages} {178005} (\bibinfo {year}
  {2020})}\BibitemShut {NoStop}%
\bibitem [{\citenamefont {Pearce}\ \emph {et~al.}(2020)\citenamefont {Pearce},
  \citenamefont {Nambisan}, \citenamefont {Ellis}, \citenamefont {Dogic},
  \citenamefont {Fernandez-Nieves},\ and\ \citenamefont {Giomi}}]{pearce2020}%
  \BibitemOpen
  \bibfield  {author} {\bibinfo {author} {\bibfnamefont {D.}~\bibnamefont
  {Pearce}}, \bibinfo {author} {\bibfnamefont {J.}~\bibnamefont {Nambisan}},
  \bibinfo {author} {\bibfnamefont {P.}~\bibnamefont {Ellis}}, \bibinfo
  {author} {\bibfnamefont {Z.}~\bibnamefont {Dogic}}, \bibinfo {author}
  {\bibfnamefont {A.}~\bibnamefont {Fernandez-Nieves}}, \ and\ \bibinfo
  {author} {\bibfnamefont {L.}~\bibnamefont {Giomi}},\ }\href@noop {}
  {\bibfield  {journal} {\bibinfo  {journal} {arXiv preprint arXiv:2004.13704}\
  } (\bibinfo {year} {2020})}\BibitemShut {NoStop}%
\bibitem [{\citenamefont {Vafa}\ \emph {et~al.}(2020)\citenamefont {Vafa},
  \citenamefont {Bowick}, \citenamefont {Marchetti},\ and\ \citenamefont
  {Shraiman}}]{vafa2020}%
  \BibitemOpen
  \bibfield  {author} {\bibinfo {author} {\bibfnamefont {F.}~\bibnamefont
  {Vafa}}, \bibinfo {author} {\bibfnamefont {M.~J.}\ \bibnamefont {Bowick}},
  \bibinfo {author} {\bibfnamefont {M.~C.}\ \bibnamefont {Marchetti}}, \ and\
  \bibinfo {author} {\bibfnamefont {B.~I.}\ \bibnamefont {Shraiman}},\
  }\href@noop {} {\bibfield  {journal} {\bibinfo  {journal} {arXiv preprint
  arXiv:2007.02947}\ } (\bibinfo {year} {2020})}\BibitemShut {NoStop}%
\bibitem [{\citenamefont {Serra}\ \emph {et~al.}(2021)\citenamefont {Serra},
  \citenamefont {Lemma}, \citenamefont {Giomi}, \citenamefont {Dogic},\ and\
  \citenamefont {Mahadevan}}]{serra2021}%
  \BibitemOpen
  \bibfield  {author} {\bibinfo {author} {\bibfnamefont {M.}~\bibnamefont
  {Serra}}, \bibinfo {author} {\bibfnamefont {L.}~\bibnamefont {Lemma}},
  \bibinfo {author} {\bibfnamefont {L.}~\bibnamefont {Giomi}}, \bibinfo
  {author} {\bibfnamefont {Z.}~\bibnamefont {Dogic}}, \ and\ \bibinfo {author}
  {\bibfnamefont {L.}~\bibnamefont {Mahadevan}},\ }\href@noop {} {\bibfield
  {journal} {\bibinfo  {journal} {arXiv preprint arXiv:2104.02196}\ } (\bibinfo
  {year} {2021})}\BibitemShut {NoStop}%
\bibitem [{\citenamefont {Mueller}\ and\ \citenamefont
  {Doostmohammadi}(2021)}]{mueller2021}%
  \BibitemOpen
  \bibfield  {author} {\bibinfo {author} {\bibfnamefont {R.}~\bibnamefont
  {Mueller}}\ and\ \bibinfo {author} {\bibfnamefont {A.}~\bibnamefont
  {Doostmohammadi}},\ }\href@noop {} {\bibfield  {journal} {\bibinfo  {journal}
  {arXiv preprint arXiv:2102.05557}\ } (\bibinfo {year} {2021})}\BibitemShut
  {NoStop}%
\bibitem [{\citenamefont {Mahault}\ and\ \citenamefont
  {Chat{\'e}}(2021)}]{mahault2021}%
  \BibitemOpen
  \bibfield  {author} {\bibinfo {author} {\bibfnamefont {B.}~\bibnamefont
  {Mahault}}\ and\ \bibinfo {author} {\bibfnamefont {H.}~\bibnamefont
  {Chat{\'e}}},\ }\href@noop {} {\bibfield  {journal} {\bibinfo  {journal}
  {arXiv preprint arXiv:2104.05453}\ } (\bibinfo {year} {2021})}\BibitemShut
  {NoStop}%
\bibitem [{\citenamefont {Mart{\'\i}nez-Prat}\ \emph
  {et~al.}(2019)\citenamefont {Mart{\'\i}nez-Prat}, \citenamefont
  {Ign{\'e}s-Mullol}, \citenamefont {Casademunt},\ and\ \citenamefont
  {Sagu{\'e}s}}]{martinez2019}%
  \BibitemOpen
  \bibfield  {author} {\bibinfo {author} {\bibfnamefont {B.}~\bibnamefont
  {Mart{\'\i}nez-Prat}}, \bibinfo {author} {\bibfnamefont {J.}~\bibnamefont
  {Ign{\'e}s-Mullol}}, \bibinfo {author} {\bibfnamefont {J.}~\bibnamefont
  {Casademunt}}, \ and\ \bibinfo {author} {\bibfnamefont {F.}~\bibnamefont
  {Sagu{\'e}s}},\ }\href {\doibase https://doi.org/10.1038/s41567-018-0411-6}
  {\bibfield  {journal} {\bibinfo  {journal} {Nat. Phys.}\ }\textbf {\bibinfo
  {volume} {15}},\ \bibinfo {pages} {362} (\bibinfo {year} {2019})}\BibitemShut
  {NoStop}%
\bibitem [{\citenamefont {Sokolov}\ \emph {et~al.}(2019)\citenamefont
  {Sokolov}, \citenamefont {Mozaffari}, \citenamefont {Zhang}, \citenamefont
  {de~Pablo},\ and\ \citenamefont {Snezhko}}]{sokolov2019}%
  \BibitemOpen
  \bibfield  {author} {\bibinfo {author} {\bibfnamefont {A.}~\bibnamefont
  {Sokolov}}, \bibinfo {author} {\bibfnamefont {A.}~\bibnamefont {Mozaffari}},
  \bibinfo {author} {\bibfnamefont {R.}~\bibnamefont {Zhang}}, \bibinfo
  {author} {\bibfnamefont {J.~J.}\ \bibnamefont {de~Pablo}}, \ and\ \bibinfo
  {author} {\bibfnamefont {A.}~\bibnamefont {Snezhko}},\ }\href {\doibase
  10.1103/PhysRevX.9.031014} {\bibfield  {journal} {\bibinfo  {journal} {Phys.
  Rev. X}\ }\textbf {\bibinfo {volume} {9}},\ \bibinfo {pages} {031014}
  (\bibinfo {year} {2019})}\BibitemShut {NoStop}%
\bibitem [{\citenamefont {Wioland}\ \emph {et~al.}(2013)\citenamefont
  {Wioland}, \citenamefont {Woodhouse}, \citenamefont {Dunkel}, \citenamefont
  {Kessler},\ and\ \citenamefont {Goldstein}}]{wioland2013}%
  \BibitemOpen
  \bibfield  {author} {\bibinfo {author} {\bibfnamefont {H.}~\bibnamefont
  {Wioland}}, \bibinfo {author} {\bibfnamefont {F.~G.}\ \bibnamefont
  {Woodhouse}}, \bibinfo {author} {\bibfnamefont {J.}~\bibnamefont {Dunkel}},
  \bibinfo {author} {\bibfnamefont {J.~O.}\ \bibnamefont {Kessler}}, \ and\
  \bibinfo {author} {\bibfnamefont {R.~E.}\ \bibnamefont {Goldstein}},\ }\href
  {\doibase 10.1103/PhysRevLett.110.268102} {\bibfield  {journal} {\bibinfo
  {journal} {Phys. Rev. Lett.}\ }\textbf {\bibinfo {volume} {110}},\ \bibinfo
  {pages} {268102} (\bibinfo {year} {2013})}\BibitemShut {NoStop}%
\bibitem [{\citenamefont {Ravnik}\ and\ \citenamefont
  {Yeomans}(2013)}]{ravnik2013}%
  \BibitemOpen
  \bibfield  {author} {\bibinfo {author} {\bibfnamefont {M.}~\bibnamefont
  {Ravnik}}\ and\ \bibinfo {author} {\bibfnamefont {J.~M.}\ \bibnamefont
  {Yeomans}},\ }\href {\doibase https://doi.org/10.1103/PhysRevLett.110.026001}
  {\bibfield  {journal} {\bibinfo  {journal} {Phys. Rev. Lett.}\ }\textbf
  {\bibinfo {volume} {110}},\ \bibinfo {pages} {026001} (\bibinfo {year}
  {2013})}\BibitemShut {NoStop}%
\bibitem [{\citenamefont {Zhang}\ \emph
  {et~al.}(2016{\natexlab{a}})\citenamefont {Zhang}, \citenamefont {Zhou},
  \citenamefont {Rahimi},\ and\ \citenamefont {de~Pablo}}]{zhang2016}%
  \BibitemOpen
  \bibfield  {author} {\bibinfo {author} {\bibfnamefont {R.}~\bibnamefont
  {Zhang}}, \bibinfo {author} {\bibfnamefont {Y.}~\bibnamefont {Zhou}},
  \bibinfo {author} {\bibfnamefont {M.}~\bibnamefont {Rahimi}}, \ and\ \bibinfo
  {author} {\bibfnamefont {J.~J.}\ \bibnamefont {de~Pablo}},\ }\href {\doibase
  https://doi.org/10.1038/ncomms13483} {\bibfield  {journal} {\bibinfo
  {journal} {Nat. Commun.}\ }\textbf {\bibinfo {volume} {7}},\ \bibinfo {pages}
  {13483} (\bibinfo {year} {2016}{\natexlab{a}})}\BibitemShut {NoStop}%
\bibitem [{\citenamefont {Khoromskaia}\ and\ \citenamefont
  {Alexander}(2017)}]{khoromskaia2017}%
  \BibitemOpen
  \bibfield  {author} {\bibinfo {author} {\bibfnamefont {D.}~\bibnamefont
  {Khoromskaia}}\ and\ \bibinfo {author} {\bibfnamefont {G.~P.}\ \bibnamefont
  {Alexander}},\ }\href {\doibase https://doi.org/10.1088/1367-2630/aa89aa}
  {\bibfield  {journal} {\bibinfo  {journal} {New J. Phys.}\ }\textbf {\bibinfo
  {volume} {19}},\ \bibinfo {pages} {103043} (\bibinfo {year}
  {2017})}\BibitemShut {NoStop}%
\bibitem [{\citenamefont {Wu}\ \emph {et~al.}(2017)\citenamefont {Wu},
  \citenamefont {Hishamunda}, \citenamefont {Chen}, \citenamefont {DeCamp},
  \citenamefont {Chang}, \citenamefont {Fern{\'a}ndez-Nieves}, \citenamefont
  {Fraden},\ and\ \citenamefont {Dogic}}]{wu2017}%
  \BibitemOpen
  \bibfield  {author} {\bibinfo {author} {\bibfnamefont {K.-T.}\ \bibnamefont
  {Wu}}, \bibinfo {author} {\bibfnamefont {J.~B.}\ \bibnamefont {Hishamunda}},
  \bibinfo {author} {\bibfnamefont {D.~T.}\ \bibnamefont {Chen}}, \bibinfo
  {author} {\bibfnamefont {S.~J.}\ \bibnamefont {DeCamp}}, \bibinfo {author}
  {\bibfnamefont {Y.-W.}\ \bibnamefont {Chang}}, \bibinfo {author}
  {\bibfnamefont {A.}~\bibnamefont {Fern{\'a}ndez-Nieves}}, \bibinfo {author}
  {\bibfnamefont {S.}~\bibnamefont {Fraden}}, \ and\ \bibinfo {author}
  {\bibfnamefont {Z.}~\bibnamefont {Dogic}},\ }\href {\doibase
  https://doi.org/10.1126/science.aal1979} {\bibfield  {journal} {\bibinfo
  {journal} {Science}\ }\textbf {\bibinfo {volume} {355}},\ \bibinfo {pages}
  {eaal1979} (\bibinfo {year} {2017})}\BibitemShut {NoStop}%
\bibitem [{\citenamefont {Ellis}\ \emph {et~al.}(2018)\citenamefont {Ellis},
  \citenamefont {Pearce}, \citenamefont {Chang}, \citenamefont {Goldsztein},
  \citenamefont {Giomi},\ and\ \citenamefont {Fernandez-Nieves}}]{ellis2018}%
  \BibitemOpen
  \bibfield  {author} {\bibinfo {author} {\bibfnamefont {P.~W.}\ \bibnamefont
  {Ellis}}, \bibinfo {author} {\bibfnamefont {D.~J.}\ \bibnamefont {Pearce}},
  \bibinfo {author} {\bibfnamefont {Y.-W.}\ \bibnamefont {Chang}}, \bibinfo
  {author} {\bibfnamefont {G.}~\bibnamefont {Goldsztein}}, \bibinfo {author}
  {\bibfnamefont {L.}~\bibnamefont {Giomi}}, \ and\ \bibinfo {author}
  {\bibfnamefont {A.}~\bibnamefont {Fernandez-Nieves}},\ }\href {\doibase
  https://doi.org/10.1038/nphys4276} {\bibfield  {journal} {\bibinfo  {journal}
  {Nat. Phys.}\ }\textbf {\bibinfo {volume} {14}},\ \bibinfo {pages} {85}
  (\bibinfo {year} {2018})}\BibitemShut {NoStop}%
\bibitem [{\citenamefont {Guillamat}\ \emph {et~al.}(2018)\citenamefont
  {Guillamat}, \citenamefont {Kos}, \citenamefont {Hardo{\"u}in}, \citenamefont
  {Ign{\'e}s-Mullol}, \citenamefont {Ravnik},\ and\ \citenamefont
  {Sagu{\'e}s}}]{guillamat2018}%
  \BibitemOpen
  \bibfield  {author} {\bibinfo {author} {\bibfnamefont {P.}~\bibnamefont
  {Guillamat}}, \bibinfo {author} {\bibfnamefont {{\v{Z}}.}~\bibnamefont
  {Kos}}, \bibinfo {author} {\bibfnamefont {J.}~\bibnamefont {Hardo{\"u}in}},
  \bibinfo {author} {\bibfnamefont {J.}~\bibnamefont {Ign{\'e}s-Mullol}},
  \bibinfo {author} {\bibfnamefont {M.}~\bibnamefont {Ravnik}}, \ and\ \bibinfo
  {author} {\bibfnamefont {F.}~\bibnamefont {Sagu{\'e}s}},\ }\href {\doibase
  10.1126/sciadv.aao1470} {\bibfield  {journal} {\bibinfo  {journal} {Sci.
  Adv.}\ }\textbf {\bibinfo {volume} {4}},\ \bibinfo {pages} {eaao1470}
  (\bibinfo {year} {2018})}\BibitemShut {NoStop}%
\bibitem [{\citenamefont {Opathalage}\ \emph {et~al.}(2019)\citenamefont
  {Opathalage}, \citenamefont {Norton}, \citenamefont {Juniper}, \citenamefont
  {Langeslay}, \citenamefont {Aghvami}, \citenamefont {Fraden},\ and\
  \citenamefont {Dogic}}]{opathalage2019}%
  \BibitemOpen
  \bibfield  {author} {\bibinfo {author} {\bibfnamefont {A.}~\bibnamefont
  {Opathalage}}, \bibinfo {author} {\bibfnamefont {M.~M.}\ \bibnamefont
  {Norton}}, \bibinfo {author} {\bibfnamefont {M.~P.}\ \bibnamefont {Juniper}},
  \bibinfo {author} {\bibfnamefont {B.}~\bibnamefont {Langeslay}}, \bibinfo
  {author} {\bibfnamefont {S.~A.}\ \bibnamefont {Aghvami}}, \bibinfo {author}
  {\bibfnamefont {S.}~\bibnamefont {Fraden}}, \ and\ \bibinfo {author}
  {\bibfnamefont {Z.}~\bibnamefont {Dogic}},\ }\href {\doibase
  https://doi.org/10.1073/pnas.1816733116} {\bibfield  {journal} {\bibinfo
  {journal} {Proc. Natl. Acad. Sci. U.S.A.}\ }\textbf {\bibinfo {volume}
  {116}},\ \bibinfo {pages} {4788} (\bibinfo {year} {2019})}\BibitemShut
  {NoStop}%
\bibitem [{\citenamefont {Hardo{\"u}in}\ \emph {et~al.}(2020)\citenamefont
  {Hardo{\"u}in}, \citenamefont {Laurent}, \citenamefont {Lopez-Leon},
  \citenamefont {Ign{\'e}s-Mullol},\ and\ \citenamefont
  {Sagu{\'e}s}}]{hardouin2020}%
  \BibitemOpen
  \bibfield  {author} {\bibinfo {author} {\bibfnamefont {J.}~\bibnamefont
  {Hardo{\"u}in}}, \bibinfo {author} {\bibfnamefont {J.}~\bibnamefont
  {Laurent}}, \bibinfo {author} {\bibfnamefont {T.}~\bibnamefont {Lopez-Leon}},
  \bibinfo {author} {\bibfnamefont {J.}~\bibnamefont {Ign{\'e}s-Mullol}}, \
  and\ \bibinfo {author} {\bibfnamefont {F.}~\bibnamefont {Sagu{\'e}s}},\
  }\href {\doibase 10.1039/D0SM00610F} {\bibfield  {journal} {\bibinfo
  {journal} {Soft Matter}\ }\textbf {\bibinfo {volume} {16}},\ \bibinfo {pages}
  {9230} (\bibinfo {year} {2020})}\BibitemShut {NoStop}%
\bibitem [{\citenamefont {Rajabi}\ \emph {et~al.}(2021)\citenamefont {Rajabi},
  \citenamefont {Baza}, \citenamefont {Turiv},\ and\ \citenamefont
  {Lavrentovich}}]{rajabi2021}%
  \BibitemOpen
  \bibfield  {author} {\bibinfo {author} {\bibfnamefont {M.}~\bibnamefont
  {Rajabi}}, \bibinfo {author} {\bibfnamefont {H.}~\bibnamefont {Baza}},
  \bibinfo {author} {\bibfnamefont {T.}~\bibnamefont {Turiv}}, \ and\ \bibinfo
  {author} {\bibfnamefont {O.~D.}\ \bibnamefont {Lavrentovich}},\ }\href
  {\doibase https://doi.org/10.1038/s41567-020-01055-5} {\bibfield  {journal}
  {\bibinfo  {journal} {Nat. Phys.}\ }\textbf {\bibinfo {volume} {17}},\
  \bibinfo {pages} {260} (\bibinfo {year} {2021})}\BibitemShut {NoStop}%
\bibitem [{\citenamefont {Norton}\ \emph {et~al.}(2018)\citenamefont {Norton},
  \citenamefont {Baskaran}, \citenamefont {Opathalage}, \citenamefont
  {Langeslay}, \citenamefont {Fraden}, \citenamefont {Baskaran},\ and\
  \citenamefont {Hagan}}]{norton2018}%
  \BibitemOpen
  \bibfield  {author} {\bibinfo {author} {\bibfnamefont {M.~M.}\ \bibnamefont
  {Norton}}, \bibinfo {author} {\bibfnamefont {A.}~\bibnamefont {Baskaran}},
  \bibinfo {author} {\bibfnamefont {A.}~\bibnamefont {Opathalage}}, \bibinfo
  {author} {\bibfnamefont {B.}~\bibnamefont {Langeslay}}, \bibinfo {author}
  {\bibfnamefont {S.}~\bibnamefont {Fraden}}, \bibinfo {author} {\bibfnamefont
  {A.}~\bibnamefont {Baskaran}}, \ and\ \bibinfo {author} {\bibfnamefont
  {M.~F.}\ \bibnamefont {Hagan}},\ }\href {\doibase 10.1103/PhysRevE.97.012702}
  {\bibfield  {journal} {\bibinfo  {journal} {Phys. Rev. E}\ }\textbf {\bibinfo
  {volume} {97}},\ \bibinfo {pages} {012702} (\bibinfo {year}
  {2018})}\BibitemShut {NoStop}%
\bibitem [{\citenamefont {Guillamat}\ \emph {et~al.}(2016)\citenamefont
  {Guillamat}, \citenamefont {Ign{\'e}s-Mullol},\ and\ \citenamefont
  {Sagu{\'e}s}}]{guillamat2016}%
  \BibitemOpen
  \bibfield  {author} {\bibinfo {author} {\bibfnamefont {P.}~\bibnamefont
  {Guillamat}}, \bibinfo {author} {\bibfnamefont {J.}~\bibnamefont
  {Ign{\'e}s-Mullol}}, \ and\ \bibinfo {author} {\bibfnamefont
  {F.}~\bibnamefont {Sagu{\'e}s}},\ }\href {\doibase
  https://doi.org/10.1073/pnas.1600339113} {\bibfield  {journal} {\bibinfo
  {journal} {Proc. Natl. Acad. Sci. U.S.A}\ }\textbf {\bibinfo {volume}
  {113}},\ \bibinfo {pages} {5498} (\bibinfo {year} {2016})}\BibitemShut
  {NoStop}%
\bibitem [{\citenamefont {Guillamat}\ \emph {et~al.}(2017)\citenamefont
  {Guillamat}, \citenamefont {Ign{\'e}s-Mullol},\ and\ \citenamefont
  {Sagu{\'e}s}}]{guillamat2017}%
  \BibitemOpen
  \bibfield  {author} {\bibinfo {author} {\bibfnamefont {P.}~\bibnamefont
  {Guillamat}}, \bibinfo {author} {\bibfnamefont {J.}~\bibnamefont
  {Ign{\'e}s-Mullol}}, \ and\ \bibinfo {author} {\bibfnamefont
  {F.}~\bibnamefont {Sagu{\'e}s}},\ }\href {\doibase
  https://doi.org/10.1038/s41467-017-00617-1} {\bibfield  {journal} {\bibinfo
  {journal} {Nat. Commun.}\ }\textbf {\bibinfo {volume} {8}},\ \bibinfo {pages}
  {564} (\bibinfo {year} {2017})}\BibitemShut {NoStop}%
\bibitem [{\citenamefont {Peng}\ \emph {et~al.}(2016)\citenamefont {Peng},
  \citenamefont {Turiv}, \citenamefont {Guo}, \citenamefont {Wei},\ and\
  \citenamefont {Lavrentovich}}]{peng2016}%
  \BibitemOpen
  \bibfield  {author} {\bibinfo {author} {\bibfnamefont {C.}~\bibnamefont
  {Peng}}, \bibinfo {author} {\bibfnamefont {T.}~\bibnamefont {Turiv}},
  \bibinfo {author} {\bibfnamefont {Y.}~\bibnamefont {Guo}}, \bibinfo {author}
  {\bibfnamefont {Q.-H.}\ \bibnamefont {Wei}}, \ and\ \bibinfo {author}
  {\bibfnamefont {O.~D.}\ \bibnamefont {Lavrentovich}},\ }\href {\doibase
  10.1126/science.aah6936} {\bibfield  {journal} {\bibinfo  {journal}
  {Science}\ }\textbf {\bibinfo {volume} {354}},\ \bibinfo {pages} {882}
  (\bibinfo {year} {2016})}\BibitemShut {NoStop}%
\bibitem [{\citenamefont {Turiv}\ \emph {et~al.}(2020)\citenamefont {Turiv},
  \citenamefont {Koizumi}, \citenamefont {Thijssen}, \citenamefont {Genkin},
  \citenamefont {Yu}, \citenamefont {Peng}, \citenamefont {Wei}, \citenamefont
  {Yeomans}, \citenamefont {Aranson}, \citenamefont {Doostmohammadi} \emph
  {et~al.}}]{turiv2020}%
  \BibitemOpen
  \bibfield  {author} {\bibinfo {author} {\bibfnamefont {T.}~\bibnamefont
  {Turiv}}, \bibinfo {author} {\bibfnamefont {R.}~\bibnamefont {Koizumi}},
  \bibinfo {author} {\bibfnamefont {K.}~\bibnamefont {Thijssen}}, \bibinfo
  {author} {\bibfnamefont {M.~M.}\ \bibnamefont {Genkin}}, \bibinfo {author}
  {\bibfnamefont {H.}~\bibnamefont {Yu}}, \bibinfo {author} {\bibfnamefont
  {C.}~\bibnamefont {Peng}}, \bibinfo {author} {\bibfnamefont {Q.-H.}\
  \bibnamefont {Wei}}, \bibinfo {author} {\bibfnamefont {J.~M.}\ \bibnamefont
  {Yeomans}}, \bibinfo {author} {\bibfnamefont {I.~S.}\ \bibnamefont
  {Aranson}}, \bibinfo {author} {\bibfnamefont {A.}~\bibnamefont
  {Doostmohammadi}},  \emph {et~al.},\ }\href {\doibase
  https://doi.org/10.1038/s41567-020-0793-0} {\bibfield  {journal} {\bibinfo
  {journal} {Nat. Phys.}\ }\textbf {\bibinfo {volume} {16}},\ \bibinfo {pages}
  {481} (\bibinfo {year} {2020})}\BibitemShut {NoStop}%
\bibitem [{\citenamefont {Lavrentovich}(2021)}]{lavrentovich2021}%
  \BibitemOpen
  \bibfield  {author} {\bibinfo {author} {\bibfnamefont {O.~D.}\ \bibnamefont
  {Lavrentovich}},\ }\href {\doibase 10.1080/21680396.2021.1919576} {\bibfield
  {journal} {\bibinfo  {journal} {Liq. Cryst. Rev.}\ ,\ \bibinfo {pages} {1}}
  (\bibinfo {year} {2021})}\BibitemShut {NoStop}%
\bibitem [{\citenamefont {Thampi}\ \emph {et~al.}(2014)\citenamefont {Thampi},
  \citenamefont {Golestanian},\ and\ \citenamefont {Yeomans}}]{thampi2014}%
  \BibitemOpen
  \bibfield  {author} {\bibinfo {author} {\bibfnamefont {S.~P.}\ \bibnamefont
  {Thampi}}, \bibinfo {author} {\bibfnamefont {R.}~\bibnamefont {Golestanian}},
  \ and\ \bibinfo {author} {\bibfnamefont {J.~M.}\ \bibnamefont {Yeomans}},\
  }\href {\doibase 10.1103/PhysRevE.90.062307} {\bibfield  {journal} {\bibinfo
  {journal} {Phys. Rev. E}\ }\textbf {\bibinfo {volume} {90}},\ \bibinfo
  {pages} {062307} (\bibinfo {year} {2014})}\BibitemShut {NoStop}%
\bibitem [{\citenamefont {Doostmohammadi}\ \emph
  {et~al.}(2016{\natexlab{b}})\citenamefont {Doostmohammadi}, \citenamefont
  {Adamer}, \citenamefont {Thampi},\ and\ \citenamefont
  {Yeomans}}]{doostmohammadi2016}%
  \BibitemOpen
  \bibfield  {author} {\bibinfo {author} {\bibfnamefont {A.}~\bibnamefont
  {Doostmohammadi}}, \bibinfo {author} {\bibfnamefont {M.~F.}\ \bibnamefont
  {Adamer}}, \bibinfo {author} {\bibfnamefont {S.~P.}\ \bibnamefont {Thampi}},
  \ and\ \bibinfo {author} {\bibfnamefont {J.~M.}\ \bibnamefont {Yeomans}},\
  }\href {\doibase https://doi.org/10.1038/ncomms10557} {\bibfield  {journal}
  {\bibinfo  {journal} {Nat. Commun.}\ }\textbf {\bibinfo {volume} {7}},\
  \bibinfo {pages} {10557} (\bibinfo {year} {2016}{\natexlab{b}})}\BibitemShut
  {NoStop}%
\bibitem [{\citenamefont {Pearce}(2019)}]{pearce2019}%
  \BibitemOpen
  \bibfield  {author} {\bibinfo {author} {\bibfnamefont {D.}~\bibnamefont
  {Pearce}},\ }\href {\doibase 10.1103/PhysRevLett.122.227801} {\bibfield
  {journal} {\bibinfo  {journal} {Phys. Rev. Lett.}\ }\textbf {\bibinfo
  {volume} {122}},\ \bibinfo {pages} {227801} (\bibinfo {year}
  {2019})}\BibitemShut {NoStop}%
\bibitem [{\citenamefont {Thijssen}\ \emph {et~al.}(2021)\citenamefont
  {Thijssen}, \citenamefont {Khaladj}, \citenamefont {Aghvami}, \citenamefont
  {Gharbi}, \citenamefont {Fraden}, \citenamefont {Yeomans}, \citenamefont
  {Hirst},\ and\ \citenamefont {Shendruk}}]{thijssen2021}%
  \BibitemOpen
  \bibfield  {author} {\bibinfo {author} {\bibfnamefont {K.}~\bibnamefont
  {Thijssen}}, \bibinfo {author} {\bibfnamefont {D.}~\bibnamefont {Khaladj}},
  \bibinfo {author} {\bibfnamefont {S.~A.}\ \bibnamefont {Aghvami}}, \bibinfo
  {author} {\bibfnamefont {M.~A.}\ \bibnamefont {Gharbi}}, \bibinfo {author}
  {\bibfnamefont {S.}~\bibnamefont {Fraden}}, \bibinfo {author} {\bibfnamefont
  {J.~M.}\ \bibnamefont {Yeomans}}, \bibinfo {author} {\bibfnamefont {L.~S.}\
  \bibnamefont {Hirst}}, \ and\ \bibinfo {author} {\bibfnamefont {T.~N.}\
  \bibnamefont {Shendruk}},\ }\href@noop {} {\bibfield  {journal} {\bibinfo
  {journal} {arXiv preprint arXiv:2102.10184}\ } (\bibinfo {year}
  {2021})}\BibitemShut {NoStop}%
\bibitem [{\citenamefont {Zhang}\ \emph
  {et~al.}(2021{\natexlab{b}})\citenamefont {Zhang}, \citenamefont {Redford},
  \citenamefont {Ruijgrok}, \citenamefont {Kumar}, \citenamefont {Mozaffari},
  \citenamefont {Zemsky}, \citenamefont {Dinner}, \citenamefont {Vitelli},
  \citenamefont {Bryant}, \citenamefont {Gardel} \emph
  {et~al.}}]{zhang2021spatio}%
  \BibitemOpen
  \bibfield  {author} {\bibinfo {author} {\bibfnamefont {R.}~\bibnamefont
  {Zhang}}, \bibinfo {author} {\bibfnamefont {S.~A.}\ \bibnamefont {Redford}},
  \bibinfo {author} {\bibfnamefont {P.~V.}\ \bibnamefont {Ruijgrok}}, \bibinfo
  {author} {\bibfnamefont {N.}~\bibnamefont {Kumar}}, \bibinfo {author}
  {\bibfnamefont {A.}~\bibnamefont {Mozaffari}}, \bibinfo {author}
  {\bibfnamefont {S.}~\bibnamefont {Zemsky}}, \bibinfo {author} {\bibfnamefont
  {A.~R.}\ \bibnamefont {Dinner}}, \bibinfo {author} {\bibfnamefont
  {V.}~\bibnamefont {Vitelli}}, \bibinfo {author} {\bibfnamefont
  {Z.}~\bibnamefont {Bryant}}, \bibinfo {author} {\bibfnamefont {M.~L.}\
  \bibnamefont {Gardel}},  \emph {et~al.},\ }\href {\doibase
  https://doi.org/10.1038/s41563-020-00901-4} {\bibfield  {journal} {\bibinfo
  {journal} {Nat. Mater.}\ ,\ \bibinfo {pages} {1}} (\bibinfo {year}
  {2021}{\natexlab{b}})}\BibitemShut {NoStop}%
\bibitem [{\citenamefont {Shankar}\ and\ \citenamefont
  {Marchetti}(2019)}]{shankar2019}%
  \BibitemOpen
  \bibfield  {author} {\bibinfo {author} {\bibfnamefont {S.}~\bibnamefont
  {Shankar}}\ and\ \bibinfo {author} {\bibfnamefont {M.~C.}\ \bibnamefont
  {Marchetti}},\ }\href {\doibase 10.1103/PhysRevX.9.041047} {\bibfield
  {journal} {\bibinfo  {journal} {Phys. Rev. X}\ }\textbf {\bibinfo {volume}
  {9}},\ \bibinfo {pages} {041047} (\bibinfo {year} {2019})}\BibitemShut
  {NoStop}%
\bibitem [{\citenamefont {Tang}\ and\ \citenamefont
  {Selinger}(2021)}]{tang2021}%
  \BibitemOpen
  \bibfield  {author} {\bibinfo {author} {\bibfnamefont {X.}~\bibnamefont
  {Tang}}\ and\ \bibinfo {author} {\bibfnamefont {J.~V.}\ \bibnamefont
  {Selinger}},\ }\href {\doibase 10.1103/PhysRevE.103.022703} {\bibfield
  {journal} {\bibinfo  {journal} {Phys. Rev. E}\ }\textbf {\bibinfo {volume}
  {103}},\ \bibinfo {pages} {022703} (\bibinfo {year} {2021})}\BibitemShut
  {NoStop}%
\bibitem [{\citenamefont {Zhang}\ \emph
  {et~al.}(2016{\natexlab{b}})\citenamefont {Zhang}, \citenamefont {Roberts},
  \citenamefont {Aranson},\ and\ \citenamefont {de~Pablo}}]{zhang2016jcp}%
  \BibitemOpen
  \bibfield  {author} {\bibinfo {author} {\bibfnamefont {R.}~\bibnamefont
  {Zhang}}, \bibinfo {author} {\bibfnamefont {T.}~\bibnamefont {Roberts}},
  \bibinfo {author} {\bibfnamefont {I.~S.}\ \bibnamefont {Aranson}}, \ and\
  \bibinfo {author} {\bibfnamefont {J.~J.}\ \bibnamefont {de~Pablo}},\ }\href
  {\doibase https://doi.org/10.1063/1.4940342} {\bibfield  {journal} {\bibinfo
  {journal} {J. Chem. Phys.}\ }\textbf {\bibinfo {volume} {144}},\ \bibinfo
  {pages} {084905} (\bibinfo {year} {2016}{\natexlab{b}})}\BibitemShut
  {NoStop}%
\bibitem [{\citenamefont {Zhang}\ \emph {et~al.}(2018)\citenamefont {Zhang},
  \citenamefont {Kumar}, \citenamefont {Ross}, \citenamefont {Gardel},\ and\
  \citenamefont {de~Pablo}}]{zhang2018}%
  \BibitemOpen
  \bibfield  {author} {\bibinfo {author} {\bibfnamefont {R.}~\bibnamefont
  {Zhang}}, \bibinfo {author} {\bibfnamefont {N.}~\bibnamefont {Kumar}},
  \bibinfo {author} {\bibfnamefont {J.~L.}\ \bibnamefont {Ross}}, \bibinfo
  {author} {\bibfnamefont {M.~L.}\ \bibnamefont {Gardel}}, \ and\ \bibinfo
  {author} {\bibfnamefont {J.~J.}\ \bibnamefont {de~Pablo}},\ }\href {\doibase
  https://doi.org/10.1073/pnas.1713832115} {\bibfield  {journal} {\bibinfo
  {journal} {Proc. Natl. Acad. Sci. U.S.A.}\ }\textbf {\bibinfo {volume}
  {115}},\ \bibinfo {pages} {E124} (\bibinfo {year} {2018})}\BibitemShut
  {NoStop}%
\bibitem [{\citenamefont {Kamien}(2002)}]{kamien2002}%
  \BibitemOpen
  \bibfield  {author} {\bibinfo {author} {\bibfnamefont {R.~D.}\ \bibnamefont
  {Kamien}},\ }\href {\doibase 10.1103/RevModPhys.74.953} {\bibfield  {journal}
  {\bibinfo  {journal} {Rev. Mod. Phys.}\ }\textbf {\bibinfo {volume} {74}},\
  \bibinfo {pages} {953} (\bibinfo {year} {2002})}\BibitemShut {NoStop}%
\bibitem [{\citenamefont {Vitelli}\ and\ \citenamefont
  {Nelson}(2004)}]{vitelli2004}%
  \BibitemOpen
  \bibfield  {author} {\bibinfo {author} {\bibfnamefont {V.}~\bibnamefont
  {Vitelli}}\ and\ \bibinfo {author} {\bibfnamefont {D.~R.}\ \bibnamefont
  {Nelson}},\ }\href {\doibase https://doi.org/10.1103/PhysRevE.70.051105}
  {\bibfield  {journal} {\bibinfo  {journal} {Phys. Rev. E}\ }\textbf {\bibinfo
  {volume} {70}},\ \bibinfo {pages} {051105} (\bibinfo {year}
  {2004})}\BibitemShut {NoStop}%
\bibitem [{\citenamefont {Voituriez}\ \emph {et~al.}(2005)\citenamefont
  {Voituriez}, \citenamefont {Joanny},\ and\ \citenamefont
  {Prost}}]{voituriez2005}%
  \BibitemOpen
  \bibfield  {author} {\bibinfo {author} {\bibfnamefont {R.}~\bibnamefont
  {Voituriez}}, \bibinfo {author} {\bibfnamefont {J.-F.}\ \bibnamefont
  {Joanny}}, \ and\ \bibinfo {author} {\bibfnamefont {J.}~\bibnamefont
  {Prost}},\ }\href {\doibase https://doi.org/10.1209/epl/i2004-10501-2}
  {\bibfield  {journal} {\bibinfo  {journal} {Europhys. Lett.}\ }\textbf
  {\bibinfo {volume} {70}},\ \bibinfo {pages} {404} (\bibinfo {year}
  {2005})}\BibitemShut {NoStop}%
\bibitem [{\citenamefont {Woodhouse}\ and\ \citenamefont
  {Goldstein}(2012)}]{woodhouse2012}%
  \BibitemOpen
  \bibfield  {author} {\bibinfo {author} {\bibfnamefont {F.~G.}\ \bibnamefont
  {Woodhouse}}\ and\ \bibinfo {author} {\bibfnamefont {R.~E.}\ \bibnamefont
  {Goldstein}},\ }\href {\doibase 10.1103/PhysRevLett.109.168105} {\bibfield
  {journal} {\bibinfo  {journal} {Phys. Rev. Lett.}\ }\textbf {\bibinfo
  {volume} {109}},\ \bibinfo {pages} {168105} (\bibinfo {year}
  {2012})}\BibitemShut {NoStop}%
\bibitem [{\citenamefont {Duclos}\ \emph {et~al.}(2018)\citenamefont {Duclos},
  \citenamefont {Blanch-Mercader}, \citenamefont {Yashunsky}, \citenamefont
  {Salbreux}, \citenamefont {Joanny}, \citenamefont {Prost},\ and\
  \citenamefont {Silberzan}}]{duclos2018}%
  \BibitemOpen
  \bibfield  {author} {\bibinfo {author} {\bibfnamefont {G.}~\bibnamefont
  {Duclos}}, \bibinfo {author} {\bibfnamefont {C.}~\bibnamefont
  {Blanch-Mercader}}, \bibinfo {author} {\bibfnamefont {V.}~\bibnamefont
  {Yashunsky}}, \bibinfo {author} {\bibfnamefont {G.}~\bibnamefont {Salbreux}},
  \bibinfo {author} {\bibfnamefont {J.-F.}\ \bibnamefont {Joanny}}, \bibinfo
  {author} {\bibfnamefont {J.}~\bibnamefont {Prost}}, \ and\ \bibinfo {author}
  {\bibfnamefont {P.}~\bibnamefont {Silberzan}},\ }\href {\doibase
  https://doi.org/10.1038/s41567-018-0099-7} {\bibfield  {journal} {\bibinfo
  {journal} {Nat. Phys.}\ }\textbf {\bibinfo {volume} {14}},\ \bibinfo {pages}
  {728} (\bibinfo {year} {2018})}\BibitemShut {NoStop}%
\bibitem [{\citenamefont {Segerer}\ \emph {et~al.}(2015)\citenamefont
  {Segerer}, \citenamefont {Th{\"u}roff}, \citenamefont {Alberola},
  \citenamefont {Frey},\ and\ \citenamefont {R{\"a}dler}}]{segerer2015}%
  \BibitemOpen
  \bibfield  {author} {\bibinfo {author} {\bibfnamefont {F.~J.}\ \bibnamefont
  {Segerer}}, \bibinfo {author} {\bibfnamefont {F.}~\bibnamefont
  {Th{\"u}roff}}, \bibinfo {author} {\bibfnamefont {A.~P.}\ \bibnamefont
  {Alberola}}, \bibinfo {author} {\bibfnamefont {E.}~\bibnamefont {Frey}}, \
  and\ \bibinfo {author} {\bibfnamefont {J.~O.}\ \bibnamefont {R{\"a}dler}},\
  }\href {\doibase 10.1103/PhysRevLett.114.228102} {\bibfield  {journal}
  {\bibinfo  {journal} {Phys. Rev. Lett.}\ }\textbf {\bibinfo {volume} {114}},\
  \bibinfo {pages} {228102} (\bibinfo {year} {2015})}\BibitemShut {NoStop}%
\bibitem [{\citenamefont {Liu}\ \emph {et~al.}(2021)\citenamefont {Liu},
  \citenamefont {Shankar}, \citenamefont {Marchetti},\ and\ \citenamefont
  {Wu}}]{liu2021}%
  \BibitemOpen
  \bibfield  {author} {\bibinfo {author} {\bibfnamefont {S.}~\bibnamefont
  {Liu}}, \bibinfo {author} {\bibfnamefont {S.}~\bibnamefont {Shankar}},
  \bibinfo {author} {\bibfnamefont {M.~C.}\ \bibnamefont {Marchetti}}, \ and\
  \bibinfo {author} {\bibfnamefont {Y.}~\bibnamefont {Wu}},\ }\href {\doibase
  https://doi.org/10.1038/s41586-020-03168-6} {\bibfield  {journal} {\bibinfo
  {journal} {Nature}\ }\textbf {\bibinfo {volume} {590}},\ \bibinfo {pages}
  {80} (\bibinfo {year} {2021})}\BibitemShut {NoStop}%
\bibitem [{\citenamefont {Vromans}\ and\ \citenamefont
  {Giomi}(2016)}]{vromans2016}%
  \BibitemOpen
  \bibfield  {author} {\bibinfo {author} {\bibfnamefont {A.~J.}\ \bibnamefont
  {Vromans}}\ and\ \bibinfo {author} {\bibfnamefont {L.}~\bibnamefont
  {Giomi}},\ }\href {\doibase https://doi.org/10.1039/C6SM01146B} {\bibfield
  {journal} {\bibinfo  {journal} {Soft Matter}\ }\textbf {\bibinfo {volume}
  {12}},\ \bibinfo {pages} {6490} (\bibinfo {year} {2016})}\BibitemShut
  {NoStop}%
\end{thebibliography}

%

\end{document}


\title{Defect Spirograph: Dynamical Behavior of Defects in Spatially
Patterned Active Nematics \\
Supplementary Information
}

\author{Ali Mozaffari$^{1}$}
\thanks{equal contribution}
\author{Rui Zhang$^{1,2}$}
\thanks{equal contribution}
\author{Noe Atzin$^{1}$}
\author{Juan J. de Pablo$^{1,3}$}
\email{depablo@uchicago.edu}

\affiliation{
 $^1$Pritzker School of Molecular Engineering, The University of Chicago, Chicago, Illinois
 60637, USA \\
 $^2$Current address: Department of Physics, Hong Kong University of Science and
  Technology, Clear Water Bay, Kowloon, Hong Kong\\
 $^3$Center for Molecular Engineering, Argonne National Laboratory, Lemont, Illinois
 60439, USA
}

\date{\today}

\maketitle


\section{Governing equations and numerical implementation}
The systems considered here are described in terms of a nematic tensorial order parameter
$\bf{Q}$ and a flow field $\bf{u}$. The standard theory of active nematodynamics is used to
quantify their spatiotemporal evolution.
For uniaxial systems, the nematic order parameter is written in the form ${\bf{Q}} =
S({\bf{nn}} - {\bf{I}}/3)$ where unit vector $\bf{n}$ is the nematic director field and $S$
is an order parameter that measures the extent of orientational ordering. The evolution of
this non-conserved order parameter obeys strongly non-linear equation \cite{beris1994}
\begin{equation}
 (\frac{\partial }{{\partial t}} + {\bf{u}} \cdot {\bm{\nabla}} ){\bf{Q}} - {\bf{S}} = \Gamma
 {\bf{H}},
 \label{beris}
\end{equation}
where the advection term is generalized by:
\begin{equation}
{\bf{S}} = (\xi {\bf{A}} + {\bf{\Omega }}) \cdot ({\bf{Q}} + \frac{{\bf{I}}}{3}) +
({\bf{Q}} + \frac{{\bf{I}}}{3}) \cdot (\xi {\bf{A}} - {\bf{\Omega }})
-2\xi ({\bf{Q}} + \frac{{\bf{I}}}{3})({\bf{Q}}:{\bm{\nabla}} {\bf{u}}),
\label{advection}
\end{equation}
which accounts for the response of the nematic order parameter to the symmetric ${\bf{A}}$,
and antisymmetric ${\bf{\Omega}}$, parts of velocity gradient tensor (${\bm{\nabla}}
{\bf{u}}$). Here $\xi$ is the flow aligning parameter, chosen to be $\xi = 0.7$ for flow
aligning elongated units.
The molecular field 
\begin{equation}
{\bf{H}} =  - (\frac{{\delta {\mathcal{F}_{LdG}}}}{{\delta {\bf{Q}}}} -
\frac{{\bf{I}}}{3}\Tr\frac{{\delta {\mathcal{F}_{LdG}}}}{{\delta {\bf{Q}}}}),
\label{molecular_field}
\end{equation}  
embodies the
relaxational dynamics of the nematic, which drives the system toward the minimum energy
configuration with Landau-de Gennes free energy density:
\begin{equation}
{f_{LdG}}\; = \; \frac{{{A_0}}}{2}(1 -
\frac{U}{3})\;\Tr({{\bf{Q}}^2})\; - \;\frac{{{A_0}U}}{3}\;\Tr({{\bf{Q}}^3})\;
+ \frac{{{A_0}U}}{4}\;{(\Tr({{\bf{Q}}^2}))^2} + \frac{L}{2}{({\bm{\nabla}} {\bf{Q}})^2}.
\label{free_energy}
\end{equation}
The relaxation rate is controlled by the collective rotational diffusion constant $\Gamma$.
The phenomenological coefficient $A_0$ sets the energy scale, $U$ controls the
magnitude of the order parameter, and $L$ is the elastic constant in the one-elastic
constant approximation. \par 

The local fluid density $\rho$ and velocity $\bf{u}$ are governed by the generalized
incompressible Navier-Stokes equations, modified by a frictional dissipative term
\begin{align}
 {\bm{\nabla}}  \cdot {\bf{u}} &= 0, \nonumber \\ 
 \rho (\frac{\partial }{{\partial t}} + {\bf{u}} \cdot {\bm{\nabla}} ){\bf{u}} & =
 {\bm{\nabla}}  \cdot {\bm{\Pi}} - f{\bf{u}}.
 \label{navier2}
\end{align}
The total asymmetric stress tensor
 \begin{equation}
 {\bf \Pi}={\bf \Pi} ^p+{\bf \Pi} ^a,
 \label{total_stress}
 \end{equation}
consists of the sum of a passive and an active stress, and $f$ is the friction coefficient between
the nematic fluid and the underlying substrate. The viscoelastic properties of the nematic
are lumped in the passive stress, given by the sum of viscous and elastic terms:
\begin{equation}
{{\bm{\Pi}} ^p}  = 2\eta {\bf{A}} - P_0{\bf{I}} + 2\xi ({\bf{Q}} +
\frac{{\bf{I}}}{3})({\bf{Q}}:{\bf{H}}) - \xi {\bf{H}} \cdot ({\bf{Q}} + \frac{{\bf{I}}}{3})
- \xi ({\bf{Q}} + \frac{{\bf{I}}}{3}) \cdot {\bf{H}} - {\bm{\nabla}} {\bf{Q}}:\frac{{\delta
 {\mathcal{F}_{LdG}}}}{{\delta {\bm{\nabla}} {\bf{Q}}}} + {\bf{Q}} \cdot {\bf{H}} - {\bf{H}} \cdot
 {\bf{Q}}.
 \label{viscous_stress}
 \end{equation}
The active stress is given by
\begin{equation} 
{{\bm{\Pi}} ^a} =  - \zeta {\bf{Q}}.
\label{active_stress} 
\end{equation}
Here, $\eta$ is the isotropic viscosity, $P_0$ is the isotropic bulk pressure, and $\zeta$
measures the activity strength. A flow is generated when $\bf{Q}$ experiences a spatial gradient with
$\zeta > 0$ for extensile systems and $\zeta < 0$ for contractile ones.
We employ a hybrid lattice Boltzmann method to solve the coupled governing partial
differential equations (Eqs.~\ref{beris}, \ref{navier2}) \cite{marenduzzo2007, ravnik2013,
zhang2016jcp, zhang2018, shechter2020}.
The time integration is performed
using an Euler forward scheme; the spatial derivatives are
carried out using a second order central difference and the coupling is enforced by
exchanging local fields between these algorithms at each time step.
Simulations were performed on a $150 \times 150$ two-dimensional lattice confined in a disk
of radius $R_o=75$. The medium viscosity was set to $\eta = 1/6$, and the collective
rotational viscosity was set to $\Gamma = 0.3$. We chose the following parameters throughout
the simulation $A_0 = 0.01$, $L = 0.01$, $U = 3.0$ (giving $S=0.5$), homeotropic anchoring
conditions with strength $\mathcal{W} = 0.01$, and a no-slip velocity field at the
surface of the solid disk. The system was initialized with the director field radially
oriented. The regions inside the circular domain of radius $R=25$ were activated by applying
uniform extensile active stresses to the nematic. The domain outside the diffusive
active/passive interface ($\tanh$ profile), of width $2$ lattice units, is passive.

\section{The role of active/passive boundary as a soft interface in defects dynamics}
In what follows we show how the \textit{soft}, permeable interface between active and passive regions created by the spatial patterning of activity is fundamentally different from the boundary created by a hard impenetrable wall.
In particular, we highlight the effect of active forces, ${\bf{f}}^a$, which appear at the interface between active and passive domains.
The major difference between a hard boundary and a soft interface resides in the fact that for the latter case the contributions of activity appear
as a bulk term, without need for external enforcement of anchoring. At a
soft interface, the director field adopts an orientation that is consistent with the minimization of the bulk free energy and the active flows that travel through the interface.
In contrast, a hard boundary imposes a no slip boundary condition with zero
velocity normal and tangential to the confining walls.

\begin{figure*}[!htb]
 \centering
 \includegraphics[width=0.8\textwidth]{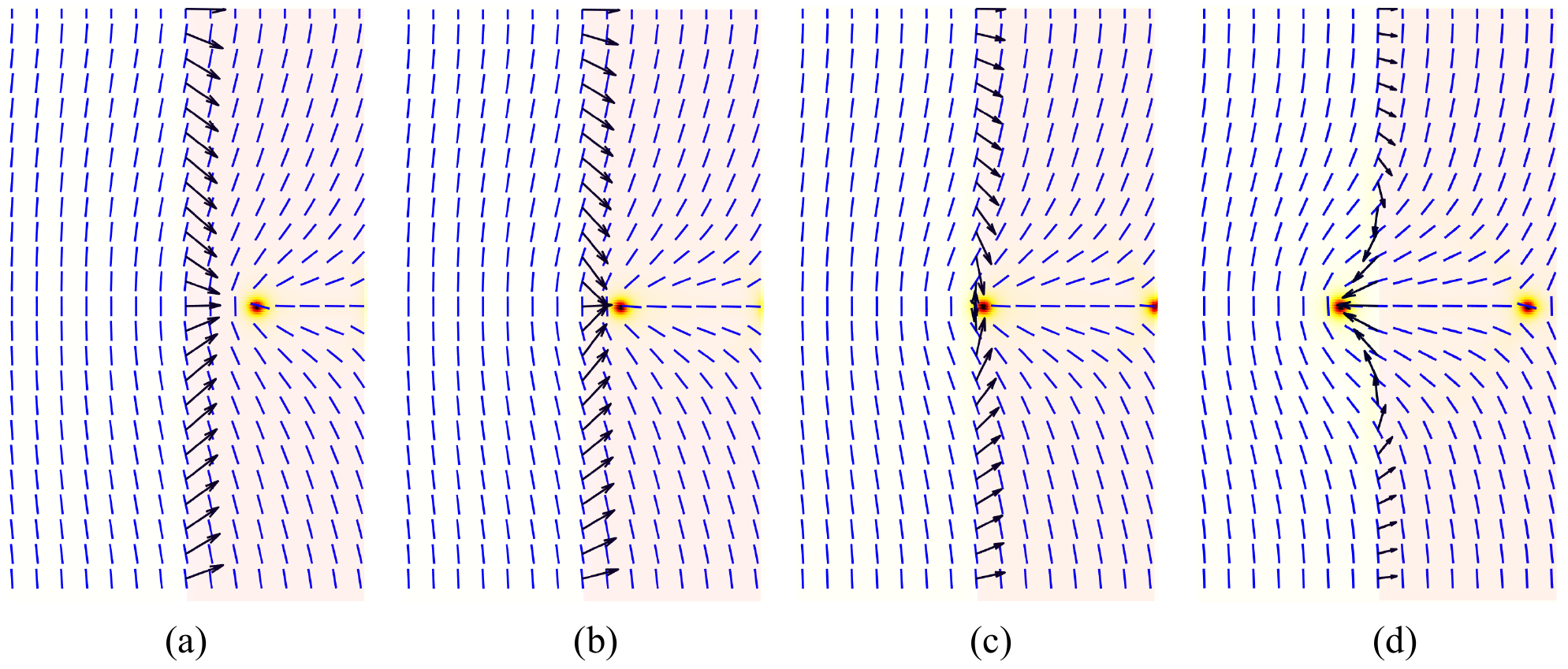}
 \caption{(a)-(d) Variation of interfacial active force as a self-propelled $+1/2$ defect approaches and crosses the interface in a perpendicular manner. The right half is active (light red) and the left one is passive. The black arrows denote the active forces which emerge at the interface, and the blue lines represent the local nematic director.}
 \label{fig_s1}
\end{figure*}

Two terms contribute to ${\bf{f}}^a$: i) the spatial
variation of the director field in the bulk of the nematic ${\bf{f}}^{a}_{\rm{b}}$, and ii) the variation in the strength
of the activity ${\bf{f}}^a_{\rm{int}}$, which is given by:
\begin{equation}
 {{\bf{f}}^a} =  {\bf{f}}^{a}_{\rm{b}} + {\bf{f}}^a_{\rm{int}} =
 - (\zeta \bm{\nabla}  \cdot {\bf{Q}} + \bm{\nabla} \zeta  \cdot {\bf{Q}}).
 \label{active_force}
\end{equation}

The bulk active force is more significant in the vicinity of defects, and decays as $1/r$ from the core of the defect.
In contrast, ${\bf{f}}^a_{\rm{int}}$ solely emerges from the activity interface. We first consider a case in which the passive
liquid crystal occupies the $x<0$ region, and the active nematic the $x>0$ region. Note that in simulations, for numerical stability
reasons, the interface is of the form $\zeta  = 1/2{\zeta _0}(\tanh (x/\omega )) + 1$, with a small
thickness $\omega$. To simplify our analysis we consider an activity pattern of the form
$\zeta  = {\zeta _0}H(x)$, where $H(x)$ is a Heaviside function. In $2$D, ${Q_{ij}} = S({n_i}{n_j} - {\delta _{ij}}/2)$, 
and the director field ${\bf{n}}({\bf{r}}) = (\cos \theta ({\bf{r}}),\sin \theta ({\bf{r}}))$, which yields the nematic order:
\begin{equation}
 {\bf{Q}} = S/2\left( {\begin{array}{*{20}{c}}
{\cos 2\theta }&{\sin 2\theta }\\
{\sin 2\theta }&{ - \cos 2\theta }
\end{array}} \right).
 \label{q_tensor}
\end{equation}
In $2$D the director field can be represented by the scalar field $\theta$, which is defined by the angle between the director and the $x$-axis in Cartesian coordinates.
It can be shown that the interfacial active force emerging right at the interface is given by:
 \begin{equation}
 {\bf{f}}_{{\mathop{\rm int}} }^a =  - {\zeta _0}\delta (x)S/2(\cos 2\theta \, {{\hat e}_x} + \sin 2\theta \,{{\hat e}_y}).
 \label{force_int}
\end{equation}
This force has a dramatic influence on the dynamics of the system. In particular, for the
case of homeotropic alignement of the director field, ($\theta = 0$), ${\bf{f}}_{{\mathop{\rm int}} }^a$ is anti-parallel to the activity gradient. A tangential anchoring of the director field, ($\theta = \pi/2$), imposes an active force along the activity gradient
(i.e. positive $x$ axis). It is worth noting that the interfacial anchoring is strongly influenced by the advective flows that are generated
by the motion of the defects, as well as the elastic forces. The frictional forces in turn stabilize the existent anchoring at the interface.

To better understand the effects of an active/passive interface we consider the following cases: \par

A) The $+1/2$ defect approaches the interface perpendicularly, driven by the active force ${\bf{f}}^{a}_{\rm{b}} = \hat p/(2r)$ 
(Fig.~\ref{fig_s1}).
Here $\hat p$ is the unit vector indicating the polarity and the axis of symmetry of the defect, which for this example is towards the
$-x$ axis. The director far field of the $+1/2$ defect induces a tangential alignment of the director field at the interface 
(corresponding to $\theta = \pi/2$ in Eq.~\ref{force_int}), resulting in
an active interfacial force along the activity gradient. Once the
self-generated flow of the defect drives the defect to the passive side, the tail of the comet-shaped defect imposes a homeotropic 
alignment at the interface (corresponding to $\theta = 0$ in Eq.~\ref{force_int}). These results clearly indicate that the
activity interface repels the $+1/2$ defect that is perpendicular to the interface
(see also \href{run:../Movies/movie_s1.m4v}{Movie S1}). \par

\begin{figure*}[!htb]
 \centering
 \includegraphics[width=0.9\textwidth]{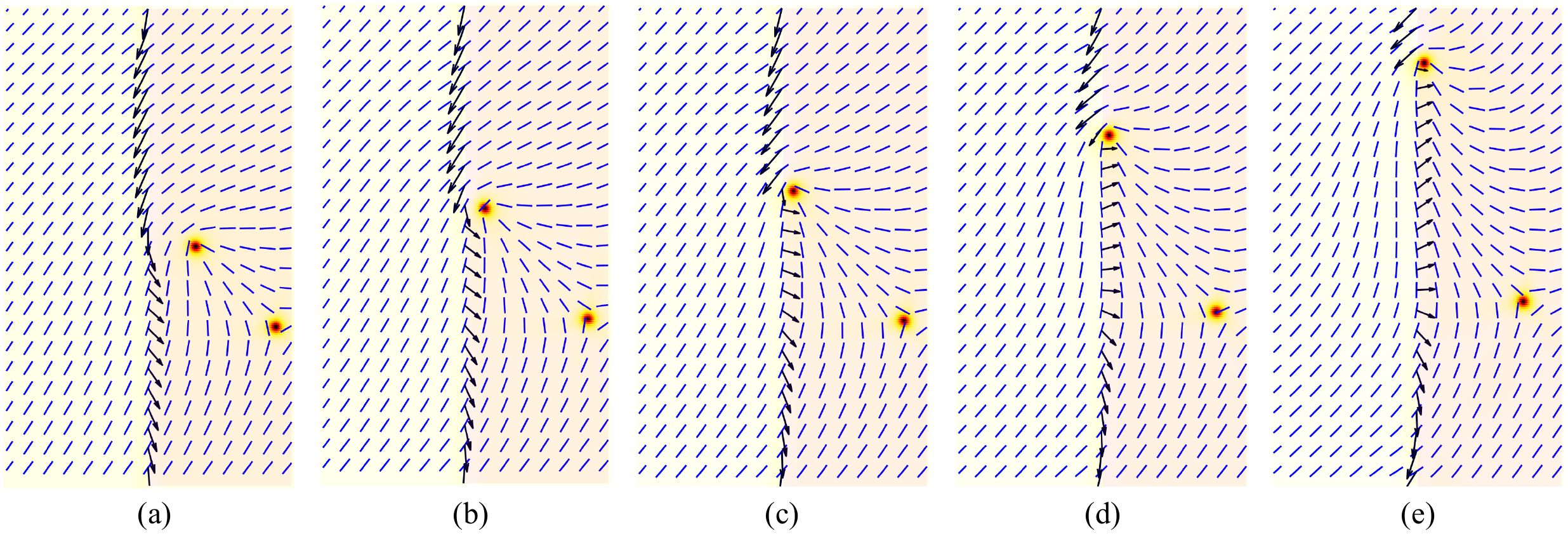}
 \caption{(a)-(e) Trajectory of $+1/2$ defect approaching an active/passive interface at an angle. The black
 arrows show the active force that arises at the activity interface, and the blue lines represent the local nematic director. The active
 domain is shaded in light red.}
 \label{fig_s2}
\end{figure*}

B) The positive half-integer charge defect approaches the interface at an angle Fig.~\ref{fig_s2}(a).
In the absence of an
activity pattern, the defect would move in a straight line. However, here the active/passive interface acts as a barrier , and adapts
dynamically to the orientation of the defect to help steer it along the interface Fig.~\ref{fig_s2}.
Note the change of director angle from
$\theta < \pi/4$ to $\theta = \pi/4$, and then to even larger values. The limit $\theta = \pi/4$ corresponds to the critical value of 
Eq.~\ref{force_int} that imposes active forces parallel to the interface (pointing anti-parallel to the defect polarity orientation) and 
hinders the penetration of the defect into the passive region
(see \href{run:../Movies/movie_s2.m4v}{Movie S2}). This mechanism helps maintain the steady rotation (SR) state observed in
Figs.~2(e),(f). The critical inclination angle of the defects above which the sliding of the defects along the interface occurs depends
strongly on the activity strength and on the frictional forces. As the activity level increases, the sliding of the defect
parallel to the interface requires the defect orientation to move further away from the interface normal. The friction in turn influences
the dynamics in an opposite way. \href{run:../Movies/movie_s3.m4v}{Movie S3} displays a case
similar to the one in Fig.~\ref{fig_s2}, but with an initial incident defect orientation angle closer to the interface normal.
The activity interface fails to prevent the penetration of the defect into the passive side, but imposes a realigning torque on the
defect that slowly reorients the defect's polarity in a direction anti-parallel to the activity gradient. \par

\begin{figure*}[!htb]
 \centering
 \includegraphics[width=0.7\textwidth]{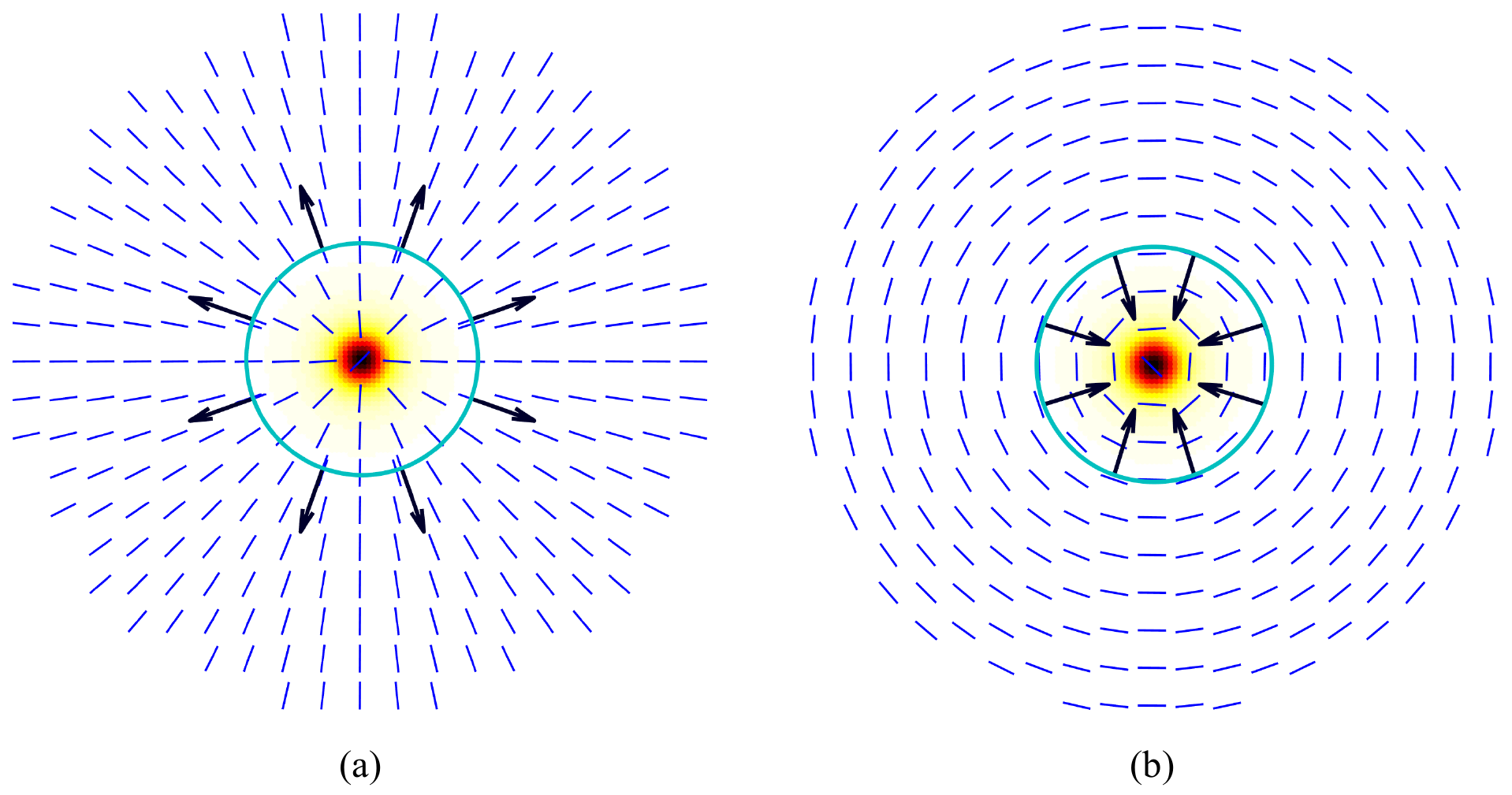}
 \caption{Active interfacial forces for a circular activity pattern. (a) Radially (b) circularly oriented director field. The blue
 circle defines the interface between the active and passive regions. }
 \label{fig_s3}
\end{figure*}

We next turn to a circular pattern of activity, which is the main focus of the current study. Simulations were performed
assuming that the regions enclosed by a circle of radius $R$ are active, with a diffuse active profile of the form
$\zeta  = {\zeta _0}(1/2 - 1/2\tanh ((\left| {{\bf{r}} - {{\bf{r}}_0}} \right| - {R})/\omega ))$. Here, $\bf{r}$ and $\bf{r}_0$ are
the position vector and the position of the system's center, respectively. For the following analysis, we assume a sharp
activity transition, with $\zeta  = {\zeta _0}(1 - H(\left| {{\bf{r}} - {{\bf{r}}_0}} \right| - {R}))$ profile. It can be shown that,
for a radial director field initialization ${\bf{n}}({\bf{r}}) = (\cos \varphi ,\sin \varphi )$, where $\phi$ is in polar coordinates, the
active interfacial force points radially outward (Fig.~\ref{fig_s3}(a)) and has the following form
 \begin{equation}
{\bf{f}}_{{\mathop{\rm int}} }^a = {\zeta _0}S/2 \, \delta (\left| {{\bf{r}} - {{\bf{r}}_0}} \right| - {R}) \,{{\hat e}_r}.
 \label{force_radial}
\end{equation}
Once the $+1$ defect at the center of the system
splits into two half-integer defects, the active forces that are strongest at the core of the defect drives them toward the activity
interface. This soft and compliant interface adopts the orientation imposed by the head of the $+1/2$ defect, and the director field switches
from radially oriented to circularly oriented. This switching leads to an interfacial active force that is directed radially towards the center, and imposes a repulsive force on the approaching defects (Fig.~\ref{fig_s3}(b)).

\section{Defect dynamics for a system with initial circular orientation of director field}
Here we probe the sensitivity of defect dynamics to the system's initial configuration. It can be shown that by initializing the director field with ${\bf{n}}({\bf{r}}) = ( - \sin \phi ,\cos \phi )$, the active interfacial force points radially inward
(Fig.~\ref{fig_s3}(b)), and can be expressed as:
\begin{equation}
{\bf{f}}_{{\mathop{\rm int}} }^a =- {\zeta _0}S/2 \, \delta (\left| {{\bf{r}} - {{\bf{r}}_0}} \right| - {R}) \,{{\hat e}_r}.
 \label{force_circular}
\end{equation}

Our analysis reveals that the initial
circular alignment leads to the expulsion of defects from the active interface. Once the $+1/2$ defects move out of the active interface, their tails enforce a normal orientation of the director field at the interface. The defects get trapped in the passive region,
not only due to the loss of activity, but also to the repulsive interfacial active forces (see Fig.~\ref{fig_s3}(a) for the direction of the
force in the case of a director normal to the interface). Interestingly, the circular configuration does not exhibit any self-organized
or coordinated defect dynamics. Besides the active turbulent state that emerges for a high level of activity, and the stagnant positive defects that are trapped at the passive region (low activity and high friction systems), other dynamics involves the generation
of additional $+1/2$ defects, which migrate out of active boundary, and the entrapment of $-1/2$ defects in the passive region
(\href{run:../Movies/movie_s4.m4v}{Movie S4}).
The depletion of $+1/2$ defects from the active region makes the system prone to bending instabilities. However, the formation of immobile $-1/2$ defects hinders the appearance of coordinated defect dynamics. This result is in stark contrast with the dynamics reported in \cite{opathalage2019, norton2018} for a hard impermeable boundary.
In \cite{norton2018}, extensile active nematics were confined by a circular boundary, with
no-slip boundary conditions and imposed tangential (and other types) anchoring, to demonstrate the insensitivity of spatiotemporal dynamics to the imposed boundary condition. Our study shows that for soft, permeable active/passive interfaces, the dynamics
are extraordinarily sensitive to the initialization of the director field; for example, the circulating state observed in \cite{opathalage2019, norton2018}
does not appear for the case of circular alignment of the director field. Additionally, \cite{opathalage2019, norton2018} discussed the accumulation of $-1/2$ defects at the solid boundary; our study shows entrapment of negative charge defects in the active region. \par
It is worth noting that, despite the strong dependence of spatiotemporal dynamics observed in our simulations on the initialization of the director field, the dynamics are independent of the anchoring strength imposed at the outer, solid boundary.   \href{run:../Movies/movie_s5.m4v}{Movie S5} displays the two-bouncing state (TB)
with no anchoring at the solid circular boundary. This results underscores the role of the active/passive soft interface in maintaining
well-organized spatiotemporal defect dynamics, and raises interesting prospects regarding its experimental relevance. \par

We also simulated a passive system with strong tangential anchoring imposed
at the outer solid boundary. Our results show anti-alignment of two $+1/2$ defects, with a center to center distance of $2 \times {5^{ - 1/4}}
R_0$, consistent with theoretical and experimental observations in \cite{duclos2017}, serving to confirm the validity of our simulations.
\par

\section{Force and torque analysis for $+1/2$ defect near an activity boundary}
 \begin{figure}[b]
 \includegraphics[scale=0.4]{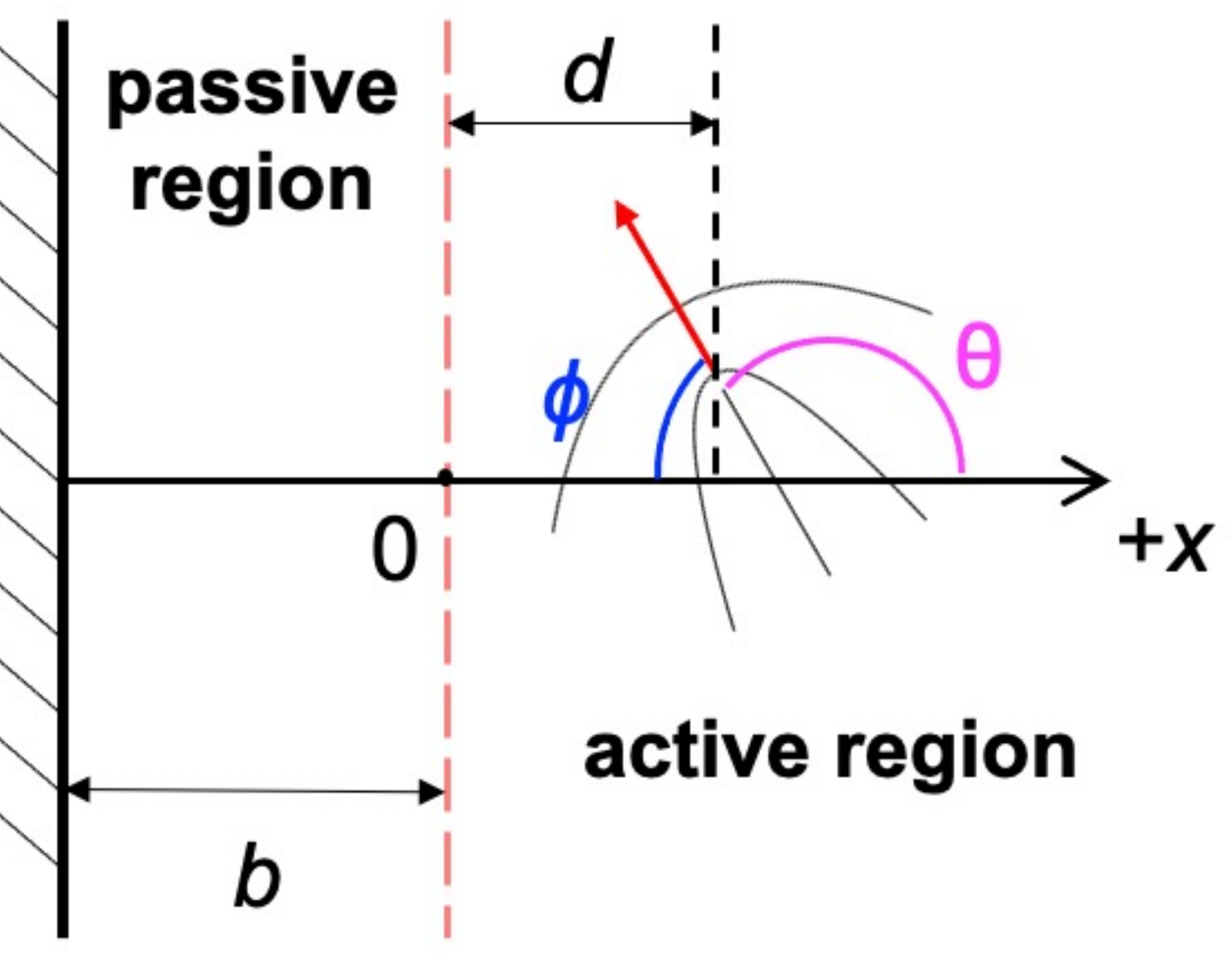}
 \caption{Schematic of the system. A +1/2 defect is near a flat active-passive interface (red dashed line) located at $x=0$. A parallel impenetrable hard wall with homeotropic anchoring is located at $x=-b$, mimicking the scenario considered in the main text by ignoring curvature effects in the circular pattern simulations. Here $\theta$ and $\phi$ denote the angles that the defect makes with the +x and -x direction, respectively.}
 \label{fig_s4}
 \end{figure}

As illustrated in Fig.~\ref{fig_s4}, we consider a +1/2 defect at $x=d$ near a flat activity interface located at the y-axis ($x=0$). A hard impenetrable wall with homeotropic anchoring conditions located at $x=-b$ is parallel to the soft interface. The defect makes an angle $\theta$ with the $+x$ direction.
The elastic energy can be written as \cite{keber2014}
$$
E^e=-\frac{\pi K}{2} \big[ \log\frac{2(d+b)}{a} + \log\cos\frac{\theta}{2}  \big],
$$
where $K$ is the elastic constant and $a$ is the defect core size. The above elastic energy captures two key physics, namely (i) the elastic energy diverges when the defect approaches the homoeotropic anchoring wall, and (ii) the $+1/2$ defect facing the $+x$ ($-x$) direction is the most (least) favorable orientation.
This elastic energy expression gives rise to the x-components of the elastic force and torque, respectively:
\begin{equation}\label{elastic}
\begin{aligned}
F_x^e &= -\frac{\partial E^e}{\partial d} = \frac{\pi K}{2} \frac{1}{d+b}; \\
T_x^e &= -\frac{\partial E^e}{\partial\theta} = -\frac{\pi K}{4} \tan \frac{\theta}{2}.
\end{aligned}
\end{equation}

To write out the active force and torque, we introduce an effective energy for a given activity level $\alpha$:
$$
E^a = \alpha h^2 (R-d) \cos\theta,
$$
where $R$ is the characteristic size of the active region, corresponding to the radius of the pattern in the circular pattern simulation. Parameter $h$ is the thickness or the hydrodynamic screening length of the quasi-2D system, determined by $\sqrt{\eta/\gamma}$, with $\eta$ and $\gamma$ being the isotropic viscosity and friction parameter, respectively. The physical picture is that only the active stress within a circular region of radius $h$ centered at the defect core can mobilize the $+1/2$ defect, due to the hydrodynamic screening provided by the friction in the system. Therefore, we can arrive at the active force and torque:
\begin{equation}\label{elastic}
\begin{aligned}
F_x^a &= \alpha h^2 \cos \theta; \\
T_x^a &=  \alpha h^2 (R-d) \sin \theta.
\end{aligned}
\end{equation}
where $F_x^a$ is simply the active force projected onto the $+x$ direction. The active torque vanishes when $\theta=0$ or $\pi$ due to symmetry reasons, and it reaches a maximum when the defect is moving parallel to the soft interface, which corresponds to the most asymmetric scenario. Such asymmetry is also distance dependent. When the defect is sufficiently far from the soft interface, the torque should vanish. Therefore, such a torque has a $(R-d)$ term.

The force and torque balance $F_x^e+F^a_x=0$ and $T^e+T^a=0$ yield
\begin{equation}\label{alpha}
\begin{aligned}
&\alpha h^2 \cos\phi = -\alpha h^2 \cos\theta =  \frac{\pi K}{2} \frac{1}{d+b}; \\
& \alpha h^2 (R-d) \sin\phi = \alpha h^2 (R-d) \sin\theta = \frac{\pi K}{4} \tan \frac{\theta}{2}.
\end{aligned}
\end{equation}
The expression above leads to the following inequality:
$$
\frac{2\sin^2\big( \frac{\phi}{2} \big)}{1-2\sin^2 \big(\frac{\phi}{2}\big) } = \frac{d+b}{2(R-d)} \ge \frac{b}{2R},
$$
therefore,
$$
2\sin^2 \big(\frac{\phi}{2} \big) \ge \frac{1}{1+2R/b}.
$$
According to Eq.~\ref{alpha}, one has
$$
\alpha = \frac{ \frac{\pi K}{4} \tan \frac{\theta}{2}}{h^2 (R-d) \sin\phi } = \frac{\pi K}{4 h^2 (R-d)\cdot 2\sin^2\big(\frac{\phi}{2}\big)} \le \frac{\pi K (1+2R/b)}{4 h^2(R-d)} \equiv \alpha_c.
$$

The above inequality implies that when activity is low, i.e. $\alpha<\alpha_c$, a solution can be found that satisfies the two balance equations, meaning that the +1/2 defect can glide along the soft interface, in which case we expect a defect rotating state for the circular patten systems. However, if the activity reaches a critical value $\alpha_c$, the solution no longer exists, indicating that the defect cannot remain close to the soft boundary, in which case we expect a defect-bouncing state. This critical activity threshold can explain the existence of a phase boundary between the defect-rotation and defect-bouncing states.
With simulation parameters $K=0.01$, $R=25$, $b=50$, $h=\sqrt{\eta/\gamma}\simeq 2$, we find $\alpha_c \sim 0.0003$, which compares favorably with the transition activity $\alpha_c\simeq 0.001$ found in full hydrodynamic simulations. \par

\begin{figure*}[!htb]
 \centering
 \includegraphics[width=0.7\textwidth]{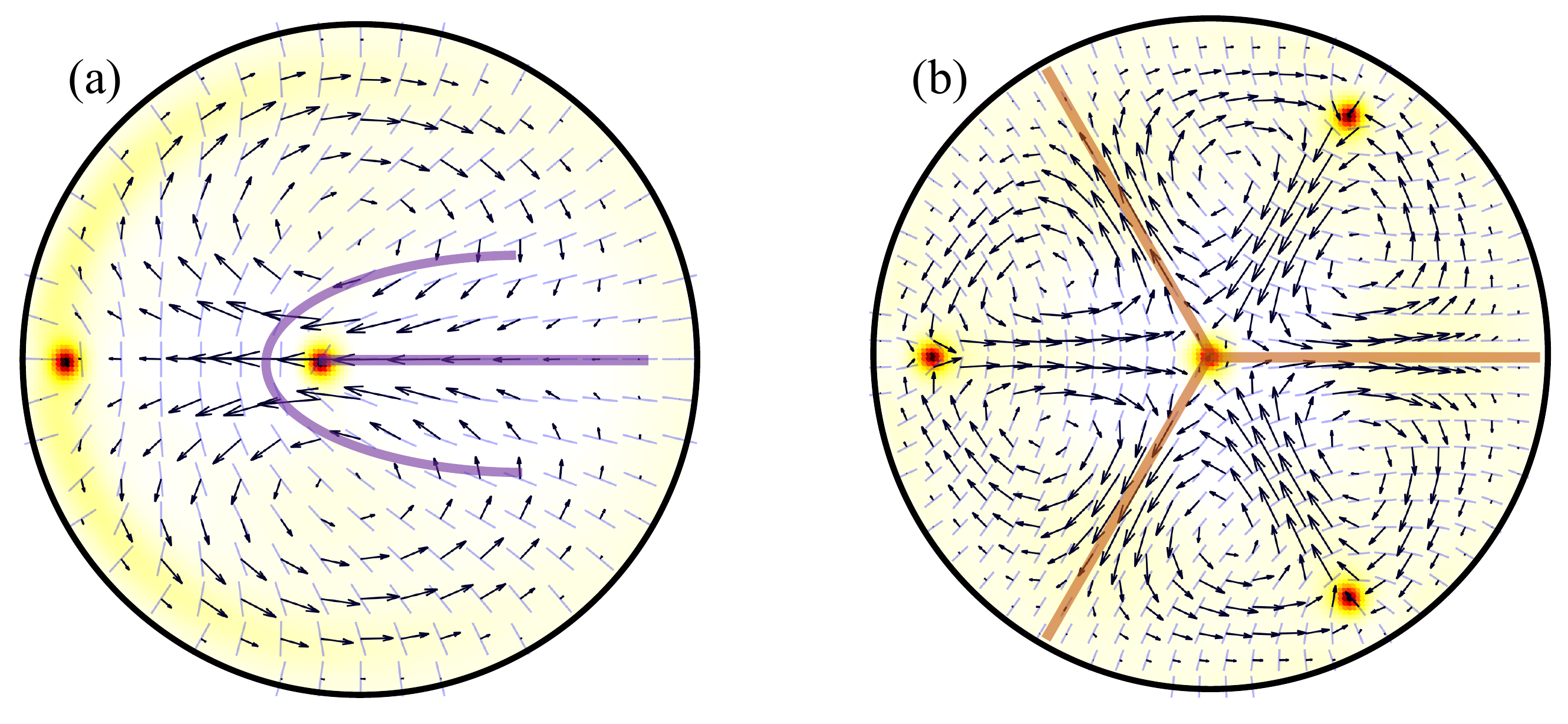}
 \caption{Self-generated flows of positive (a) and negative (b)
  half integer charge defects for a system of extensile active nematics.
  The blue lines in the background denote the director field and the black
  arrows are the velocity field. The $+1/2$ defect shows asymmetric double vortices which drive
  the defect towards its head. The $-1/2$ defect forms six-fold symmetric vortices with zero
 net force at the core of the defect. The activity is uniform, no-slip velocity and
 strong homeotropic anchoring are imposed at the outer surface of the disk. The director field is initialized
 with the ansatz to ensure the formation of the corresponding defects at the center.}
 \label{fig_s5}
\end{figure*}

\begin{figure*}[!htb]
 \centering
 \includegraphics[width=0.4\textwidth]{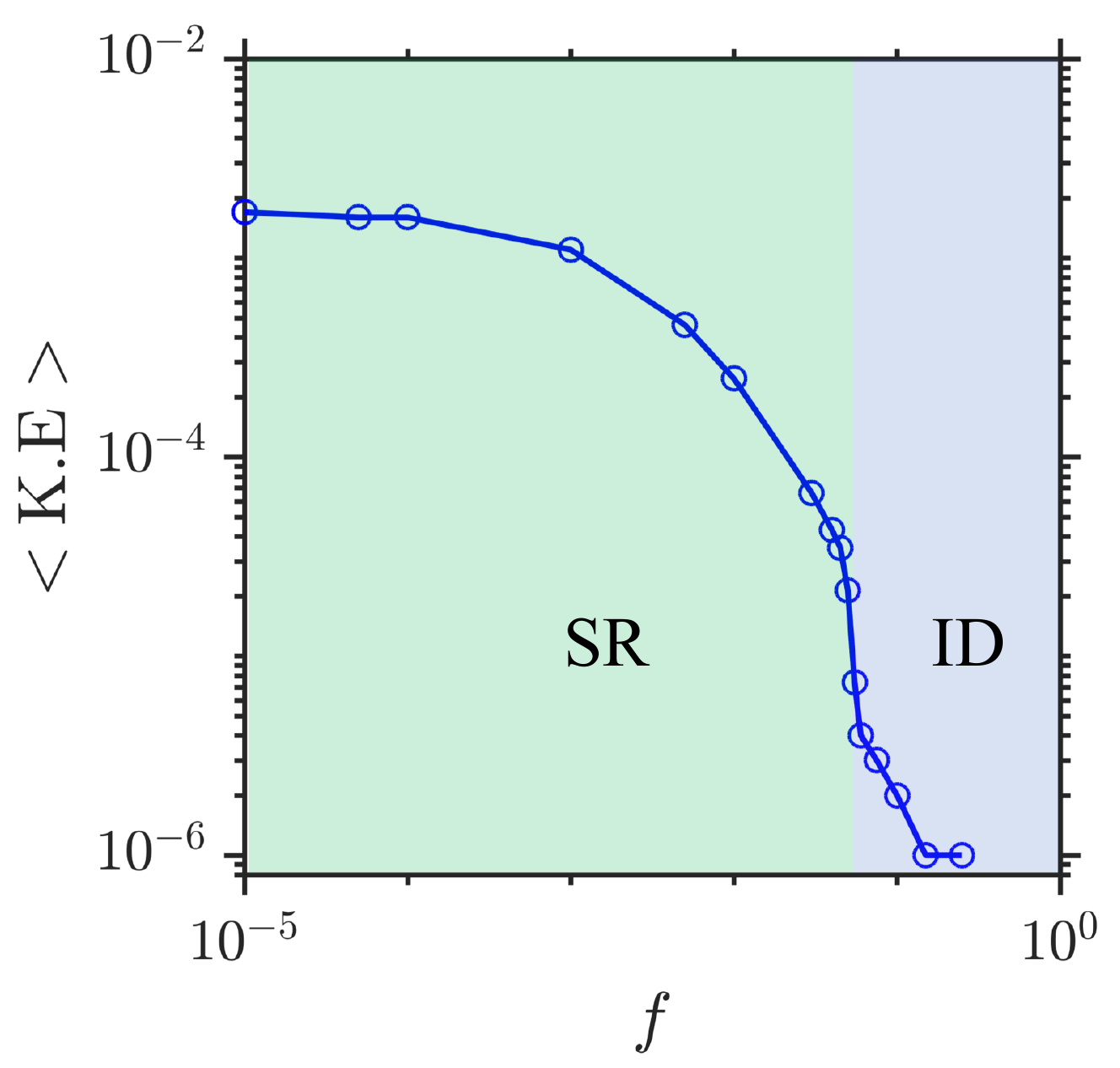}
 \caption{Effect of frictional damping forces on the average kinetic energy of the
  system. The points are associated with the blue line in Fig.~1 of the main text. The average
 kinetic energy is reduced by increasing the friction coefficient.}
 \label{fig_s6}
\end{figure*}

\clearpage


%